\documentclass[10pt,journal,compsoc]{IEEEtran}
\usepackage{amsmath,amsfonts}
\usepackage{amssymb}
\usepackage{array}

\usepackage{textcomp}
\usepackage{stfloats}
\usepackage{verbatim}
\usepackage{graphicx} %
\usepackage{cite}
\usepackage{pifont}
\usepackage[ruled,vlined]{algorithm2e}
\usepackage[justification=centering]{caption}
\usepackage{enumitem}
\usepackage{subcaption}
\usepackage{booktabs}
\usepackage{makecell}
\usepackage{todonotes}
\usepackage{url}
\usepackage{listings}
\usepackage{color}
\usepackage{tcolorbox}
\usepackage{multirow}
\usepackage{amsmath}
\usepackage{soul}
\usepackage{ragged2e}
\usepackage{threeparttable}
\usepackage{balance}
\usepackage{xcolor}
\usepackage{colortbl}
\usepackage{tabularx}
\usepackage{bbding}

\newcommand*{\new}[1]{#1}
\newcommand{\code}[1]{\texttt{#1}}

\newcommand{\grayline}{\rowcolor[gray]{.90}}

\begin{document}
\newcommand{\find}[1]{
\begin{tcolorbox}[leftrule=0.5mm,rightrule=0.5mm, toprule=0.5mm,bottomrule=0.5mm,left=2pt,right=2pt,top=2pt,bottom=2pt]%
\em #1
\end{tcolorbox}
}

\title{Enhancing Project-Specific Code Completion by Inferring Internal API Information }

\author{Le~Deng, 
        Xiaoxue~Ren, 
        Chao Ni,
        Ming Liang,
        David Lo,
        Zhongxin~Liu
\IEEEcompsocitemizethanks{\IEEEcompsocthanksitem Le Deng, Xiaoxue Ren, Chao Ni, and Zhongxin Liu are with the State Key Laboratory of Blockchain and Data Security, Zhejiang University, Hangzhou, 310027, China. \\ 
E-mail: \{dengle, xxren, chaoni, liu\_zx\}@zju.edu.cn
\IEEEcompsocthanksitem Ming Liang is with Ant Group, China. \\
Email: liangming.liang@antgroup.com
\IEEEcompsocthanksitem David Lo is with the School of Computing and Information Systems, Singapore Management University, Singapore 188065 \\
E-mail: davidlo@smu.edu.sg
\IEEEcompsocthanksitem Zhongxin Liu is the corresponding author.
}
}

\markboth{Journal of \LaTeX\ Class Files,~Vol.~14, No.~8, August~2021}%
{Shell \MakeLowercase{\textit{et al.}}: A Sample Article Using IEEEtran.cls for IEEE Journals}

\newcommand{\xiaoxue}[1]{\textcolor{orange}{\textbf{Xiaoxue}: #1}}

\IEEEtitleabstractindextext{
\begin{abstract}
\justifying{
Project-specific code completion, which aims to complete code based on the context of the project, is an important and practical software engineering task. 
The state-of-the-art approaches employ the retrieval-augmented generation (RAG) paradigm and prompt large language models (LLMs) with information retrieved from the target project for project-specific code completion.
In practice, developers always define and use custom functionalities, namely internal APIs, to facilitate the implementation of specific project requirements.
Thus, it is essential to consider internal API information for accurate project-specific code completion. 
However, existing approaches either retrieve similar code snippets, which do not necessarily contain related internal API information, or retrieve internal API information based on import statements, which usually do not exist when the related internal APIs haven't been used in the file.
Therefore, these project-specific code completion approaches face challenges in effectiveness or practicability.

To this end, this paper aims to enhance project-specific code completion by locating internal API information without relying on import statements. We first propose a method to infer internal API information.
Our method first extends the representation of each internal API by constructing its usage examples and functional semantic information (i.e., a natural language description of the function’s purpose) and constructs a knowledge base.
Based on the knowledge base, our method uses an initial completion solution generated by LLMs to infer the API information necessary for completion.
Based on this method, we propose a code completion approach that enhances project-specific code completion by integrating similar code snippets and internal API information. Furthermore, we developed a benchmark named ProjBench, which consists of recent, large-scale real-world projects and is free of leaked import statements. 
\new{We evaluated the effectiveness of our approach on ProjBench and an existing benchmark CrossCodeEval.
Experimental results show that our approach outperforms the base-performing approach by an average of +5.91 in code exact match and +6.26 in identifier exact match, corresponding to relative improvements of 22.72\% and 18.31\%, respectively. We also show our method complements existing ones by integrating it into various baselines, boosting code match by +7.77 (47.80\%) and identifier match by +8.50 (35.55\%) on average.
}
}
\end{abstract}

}

\maketitle

\section{Introduction}
Code completion is an intelligent programming task that automatically generates subsequent code based on the context of the code input by developers.
Efficient code completion can significantly reduce the programming workload by intuitively predicting and filling in necessary code, thereby boosting efficiency and minimizing errors.
Recently, a series of large language models (LLMs)~\cite{nijkamp2022codegen,li2023starcoder,lozhkov2024starcoder,guo2024deepseek,zheng2023codegeex,luo2023wizardcoder,roziere2023code} for code (i.e., code LLMs) have been proposed and demonstrate superior performance in code completion.
Some of them have been deployed as autocomplete plugins (e.g., Copilot~\cite{copilot}, CodeWhisperer~\cite{codewhisperer}) in modern IDEs, effectively enhancing developers' efficiency.

Most code LLMs are designed for independent code completion\cite{yin2018learning,chen2021evaluating}, which refers to independently generating or predicting the next pieces of code based on a given code snippet or functional description, without referencing other information.
\new{
However, in real-world software development scenarios, each project often contains specific knowledge that distinguishes it from others. We define project-specific knowledge as information that is unique to a particular codebase and not commonly found across general-purpose corpora. This usually includes the project’s internal APIs, naming conventions, and code styling practices. Such knowledge is beneficial for code completion. If large language models (LLMs) lack awareness of this knowledge, it can lead to hallucinations~\cite{zhang2023repocoder,ding2022cocomic,shrivastava2023repository} and suboptimal completions that fail to align with the project's conventions or intended functionality. Importantly, project-specific knowledge is typically not well captured during the pre-training or fine-tuning phases of LLMs, making it necessary to incorporate such information at inference time.
}

To mitigate the knowledge-lack problem, previous works~\cite{lu2022reacc,zhang2023repocoder,shrivastava2023repository,shrivastava2023repofusion,ding2022cocomic,liang2024repofuse,phan2024repohyper} typically employ the retrieval-augmented generation (RAG) paradigm~\cite{lewis2020retrieval}. For each completion task, RAG first retrieves a set of relevant code snippets from the current repository and then injects these snippets into the prompt to augment code LLMs with project-specific knowledge.
Although these methods have shown promise, they often fail to effectively capture internal API information, which refers to the unique functionalities and custom logic defined by developers according to the specific needs of a project.
\begin{figure}[t]
    \centering
    \includegraphics[width=0.48\textwidth]{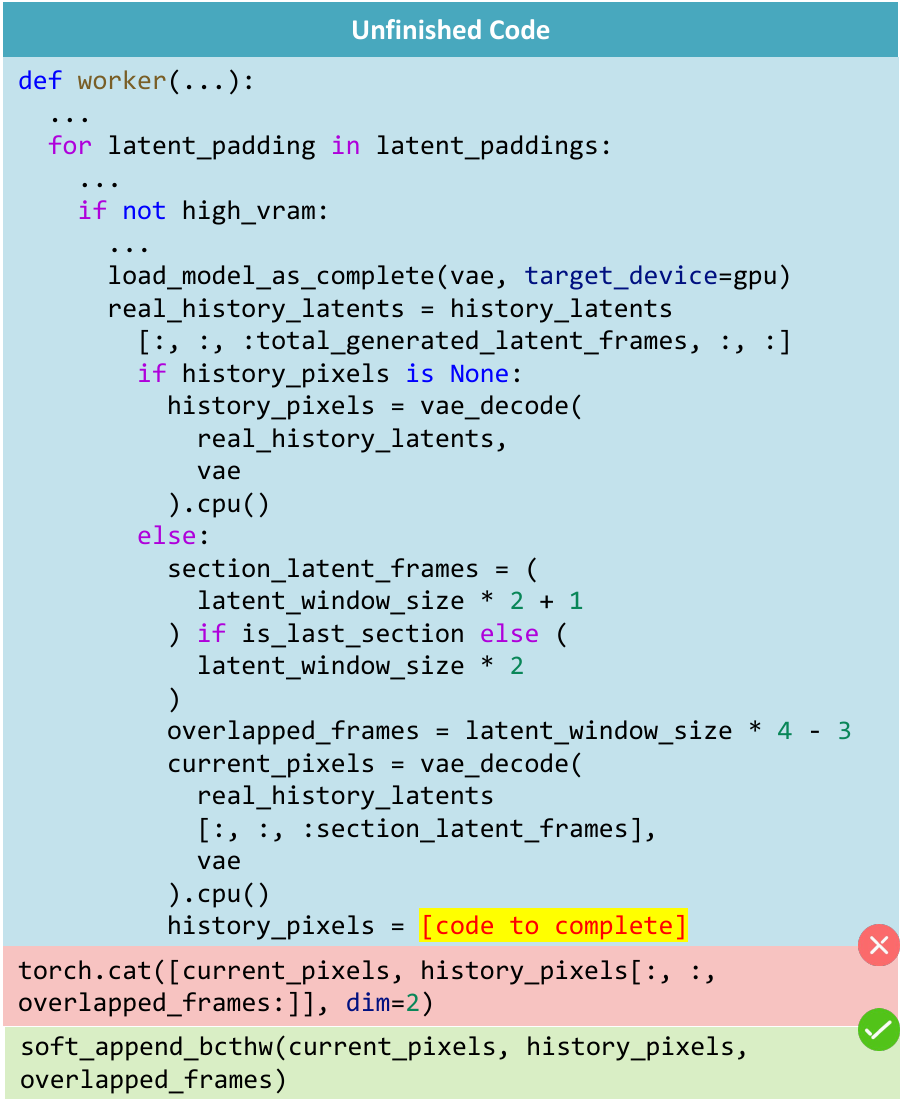}
    \caption{An example from FramePack. The red box marked with ``\XSolid'' denotes the output generated by Claude 3.7 Sonnet, while the green box marked with ``\Checkmark'' represents the ground truth.}
    \label{fig:specific_api}
\end{figure}

\new{
In contrast to standard libraries or widely-used third-party packages, whose interfaces and behaviors are well-represented in the pretraining data of code LLMs. Internal APIs are typically unseen and project-specific, making them inaccessible to LLMs without explicit in-context exposure.
In real-world scenarios, particularly in enterprise settings, these internal APIs are often proprietary or rapidly evolving and are defined and modified within the scope of individual projects. As such, they are inherently unknown to general-purpose LLMs, regardless of model scale or training data coverage.
As a result, relying solely on pretrained knowledge is insufficient for practical, project-specific code completion tasks. LLMs are likely to hallucinate or produce factually incorrect completions if such internal information is not provided.
For example, in Figure~\ref{fig:specific_api}, we present a case from FramePack (created on April 12, 2025), where we use one of the most advanced models, Claude 3.7 Sonnet (with a training cutoff of November 2024), to perform code completion. Although the model generally understands the intended functionality (i.e., concatenation), it fails to correctly generate the target code because it has never encountered the internal API \code{soft\_append\_bcthw}, which is defined internally within the project.
Therefore, to generate fact-based outputs and avoid hallucinations, it is necessary for LLMs to integrate this kind of project-specific internal API information.
}

Specifically, existing RAG-based works usually adopt similarity-based methods and dependency-based methods to retrieve relevant information, but both have limitations. Similarity-based methods~\cite{lu2022reacc,zhang2023repocoder} divide the repository into multiple code snippets and then calculate the similarity between the unfinished code and these snippets to retrieve similar ones. However, these methods \textbf{struggle to explicitly locate the internal API information} necessary for completion. Specifically, when completing internal API calls, the repository may not contain snippets that use the API. Or due to the complexity of the API's parameter list, it may be impossible to find snippets that exactly match the expected usage. Code LLMs lack a deep understanding of project internal APIs, which can result in hallucinations. Dependency-based methods retrieve more comprehensive project knowledge by acquiring the overall structure of the project and cross-file dependencies~\cite{shrivastava2023repository,shrivastava2023repofusion,ding2022cocomic,liang2024repofuse,phan2024repohyper}. These methods mainly leverage import relationships between files to obtain the cross-file information that the unfinished code depends on. Nonetheless, this approach \textbf{deviates from practical application scenarios}. Specifically, before an internal API is used for the first time within a project, its corresponding import statement may not exist, making these methods prone to failure in real-world application~\cite{semenkin2024full}.
 
In this paper, we follow the RAG paradigm while exploring a more practical method for retrieving internal API information and a project-specific code completion approach. We first propose an internal API inference method, which assists retrieval by mining the latent information of APIs and utilizing an initial completion solution generated by LLMs based on the unfinished code (referred to as code draft). Specifically, our API inference method first extends the representation of each internal API by constructing its usage example and functional semantic information (natural language descriptions of the functionality implemented by a piece of code), and builds an API knowledge base. Then, it uses a code LLM to generate a code draft. Finally, it extracts the API invocation code and functional semantic description from the code draft to retrieve the API information required for code completion from the knowledge base. Our API inference method can be used independently to supplement API information for existing RAG-based methods. 

Based on the API inference method, we further propose a novel project-specific code completion framework that not only considers similar code but also captures the internal API information on which the completion depends. Specifically, first, our approach concatenates similar code snippets with the unfinished code and inputs them into a code LLM to generate a code draft. Then, using this code draft, our approach performs further API retrieval and similar code retrieval. Finally, it combines all the retrieved project knowledge to construct a new prompt and uses the LLM to generate the target code. 

\new{
Our approach explicitly retrieves the internal API information essential for accurate code completion, thereby overcoming a key limitation of similarity-based methods, which often fail to capture the semantic relevance of internal APIs. 
In contrast to dependency-based methods that rely on import statements, our approach does not depend on explicit declarations. 
Import statements are frequently incomplete or unavailable in real-world, in-progress development scenarios. 
While we adopt a preliminary code draft to provide contextual signals, we find that relying solely on it is insufficient for retrieving the internal API knowledge required for high-quality code completion. 
Although the code draft contains rich information, it is often inaccurate or fails to include the appropriate internal API calls, especially in complex or unfamiliar scenarios. This makes it challenging to extract the truly relevant cues for retrieval. 
Therefore, how to effectively interpret and leverage such imperfect yet informative drafts becomes a crucial problem. 
To this end, we propose the Usage Example Retrieval and Functional Semantic Retrieval components to address it explicitly. 
They go beyond surface similarity by reasoning about the likely API usage and retrieving semantically relevant information. 
These components form the core of our technical contribution, enabling robust, inference-driven integration of internal project knowledge and setting our approach apart from prior retrieval paradigms.
}

Previous benchmarks~\cite{liu2023repobench,shrivastava2023repository,shrivastava2023repofusion,ding2022cocomic,ding2024crosscodeeval, zhang2023repocoder} for project-specific code completion have issues such as data leakage, lack of representativeness, or misalignment with typical usage. To better evaluate the effectiveness of project-specific code completion approaches, we construct a new benchmark named ProjBench. ProjBench is constructed from large Python and Java repositories. To avoid data leakage, we collect the newest possible projects from GitHub. To ensure that our benchmark is sufficiently representative, we select popular projects based on the number of stars. For samples where the code lines are the first use of cross-file dependencies in the current file, we mask the corresponding import statements to mimic real-world scenarios. We compare our approach with the state-of-the-art project-specific code completion framework across multiple benchmarks, including the benchmark constructed in this work. Experimental results show that our approach outperforms the best-performing baseline by an average of +5.91 in code exact match and +6.26 in identifier exact match. We also demonstrate the complementarity of our API inference method with existing methods by integrating it with various baselines.

In summary, the contributions of this paper include:
\begin{itemize}
    \item We propose a novel method to retrieve internal API information, which can complement existing RAG-based code completion works to enhance their performance. Based on this method, we develop a project-specific code completion approach, which integrates similar code with the internal API information necessary for completion.
    \item We develop a benchmark named ProjBench for project-specific code completion tasks. This benchmark is closely aligned with real-world projects in Python and Java and incorporates sufficient dependencies and project context. To the best of our knowledge, we are the first to consider the issue of import statements leakage when constructing the dataset.
    \item We evaluate our approach on ProjBench and CrossCodeEval~\cite{ding2024crosscodeeval} using code match and identifier match. Experimental results demonstrate that our approach outperforms the baselines and can complement existing RAG-base methods.
\end{itemize}

\section{Motivation}\label{sec:motivation}
In this section, we illustrate the motivation of our approach through two practical examples.

\begin{figure}[t]
    \centering
    \includegraphics[width=0.48\textwidth]{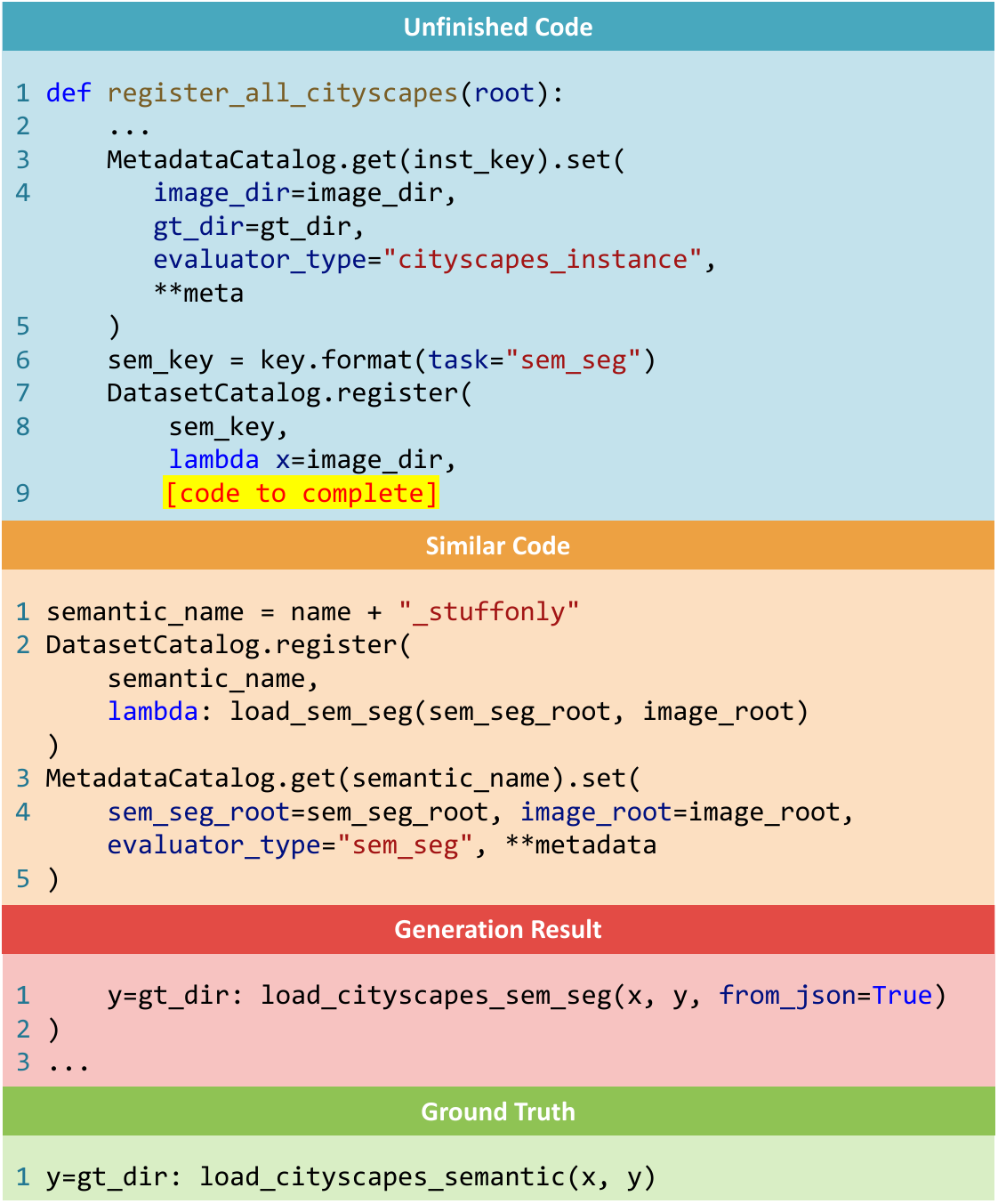}
    \caption{A example from MagicClothing}
    \label{fig:motivation_1}
\end{figure}

In Figure \ref{fig:motivation_1}, the developer is writing a function within a project\footnote{https://github.com/ShineChen1024/MagicClothing} and needs to complete the parameter list for the function call to \code{register} (blue box labeled \new{\textit{Unfinished Code}}). To complete this line of code, it is necessary to have a thorough understanding of the context of the current file as well as some APIs defined within the project~\cite{yu2024codereval,li2024deveval}.
For this example, RepoCoder retrieves a code snippet (orange box \new{labeled \textit{Similar Code}}) similar to the unfinished code and provides it to the code LLM as project knowledge. The model, by mimicking the similar snippet, generates an incorrect result (red box \new{labeled \textit{Generation Result}}), which includes a nonexistent function \code{load\_cityscapes\_sem\_seg}. RepoCoder's retrieval is based on surface-level code similarity, which fails to ensure that the LLM is provided with genuinely useful project-specific knowledge, such as which APIs within the project can be used to fulfill the current requirements. Although the retrieved snippet shares multiple tokens with the unfinished code, it does not contribute meaningfully to the correct code completion. Due to its lack of understanding of the internal APIs, RepoCoder often incorrectly uses the internal APIs, such as calling incorrect functions, generating nonexistent functions, or misusing parameters. Upon searching within the project, we find a function with a similar name to the previously incorrectly generated non-existent function, \code{load\_cityscapes\_semantic}. When we provide this function's information to the model, it correctly completes the code (green box \new{labeled \textit{Ground Truth}}).
 
\begin{figure}[t]
    \centering
    \includegraphics[width=0.48\textwidth]{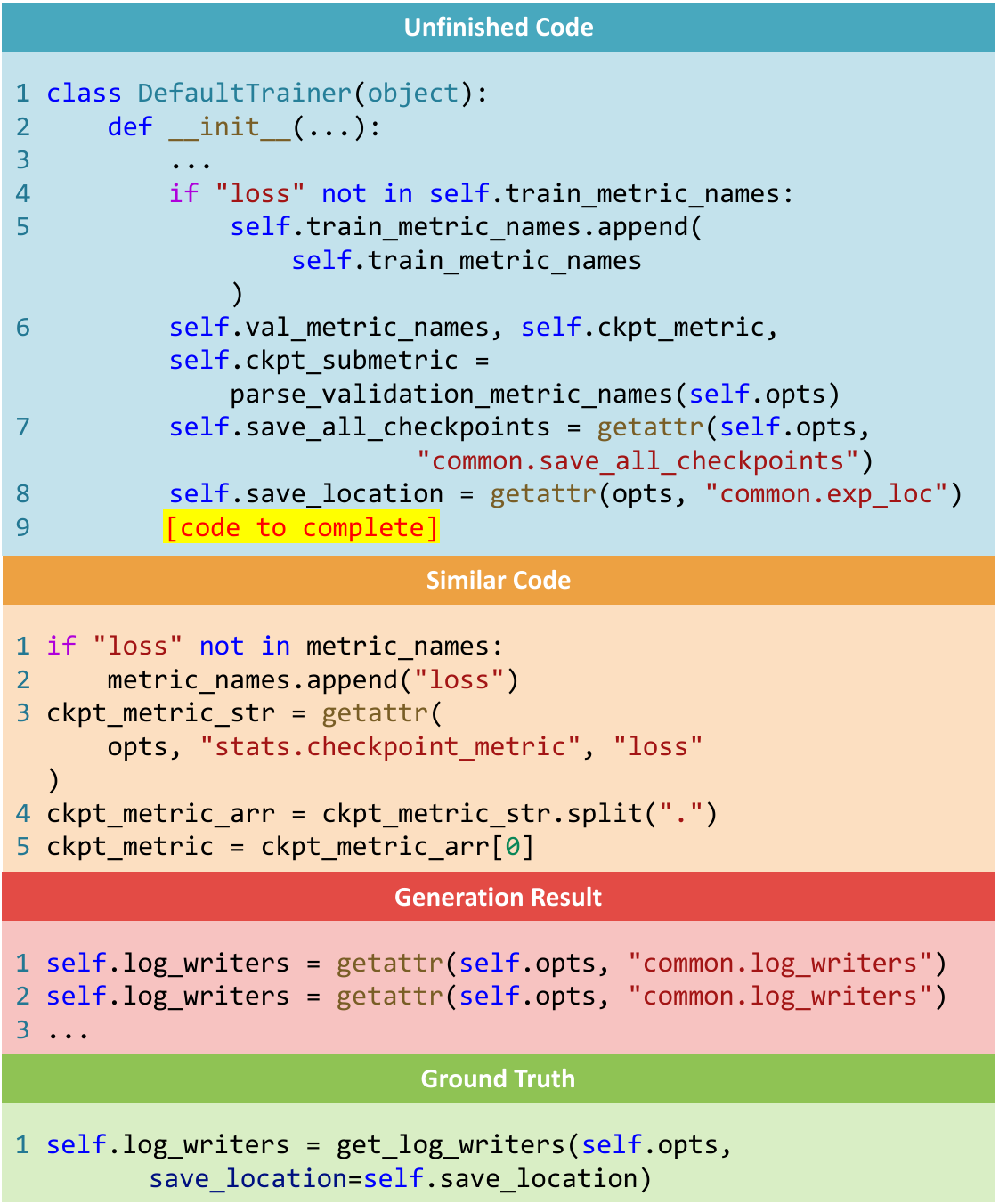}
    \caption{A example from corenet}
    \label{fig:motivation_2}
\end{figure}

In Figure \ref{fig:motivation_2}, the developer is refining the \code{\_\_init\_\_} function of \code{DefaultTrainer} within a project\footnote{https://github.com/apple/corenet} and needs to continue initializing class variables (blue box \new{\textit{labeled Unfinished Code}}).
For this example, RepoCoder searches for similar code snippets (orange box \new{labeled \textit{Similar Code}}) based on similarity metrics. It then concatenates the similar code with the unfinished code and inputs it as a prompt to the code LLM. The model, combining the similar snippet and in-file information, completes the code. However, the completion result mistakenly uses \code{getattr} to initialize \code{log\_writers} (red box \new{labeled \textit{Generation Result}}) and gets stuck in undesirable sentence-level loops~\cite{xu2022learning}. When there is no highly similar code snippet within the project, RepoCoder fails to find reusable code for reference, preventing the model from obtaining valuable cross-file information. Although the model correctly understands the completion intent based on context information (such as initializing \code{self.log\_writers}), it fails to generate useful code because it cannot identify the relevant API within the project needed to achieve this intent. This results in generating redundant and useless code, and the complexity of predicting too much code may lead to unexpected errors. Due to the lack of project-specific insights, even when RepoCoder identifies the completion intent, it cannot directly use the project's APIs to fulfill the requirements. Upon searching the entire code repository, we find a function named \code{get\_log\_writers} that can be used to obtain a list of log writers. When we provide this function's information to the model, it generates the expected result (green box \new{labeled \textit{Ground Truth}}).

RepoCoder excels at retrieving superficially similar code from the repository and can effectively reuse identical code. However, when there is no highly similar code in the project, simple copying can result in generated outcomes that deviate from actual requirements, particularly when handling unseen project-specific APIs. For example, it might incorrectly fill in function parameters, call non-existent functions, or write extensive code with similar functionality instead of calling an existing function defined within the project. Relying solely on the surface-level similarity between code snippets fails to effectively capture the critical internal API information required for accurate completion. Although some works~\cite{liang2024repofuse,phan2024repohyper} have recognized this issue, they obtain internal API information by using import statements that may not be present, which can lead to failures in real-world scenarios.

Considering the limitations of existing work, we aim to retrieve necessary internal API information for completion without relying on import statements. 
Motivated by the examples mentioned above, we find that if we extract incorrect API calls and semantic information from the initial completion results of the code LLMs to guide us in retrieving API information, it is possible to achieve this goal.

\section{Approach}
\begin{figure*}[t]
    \centering
    \includegraphics[width=1.0\textwidth]{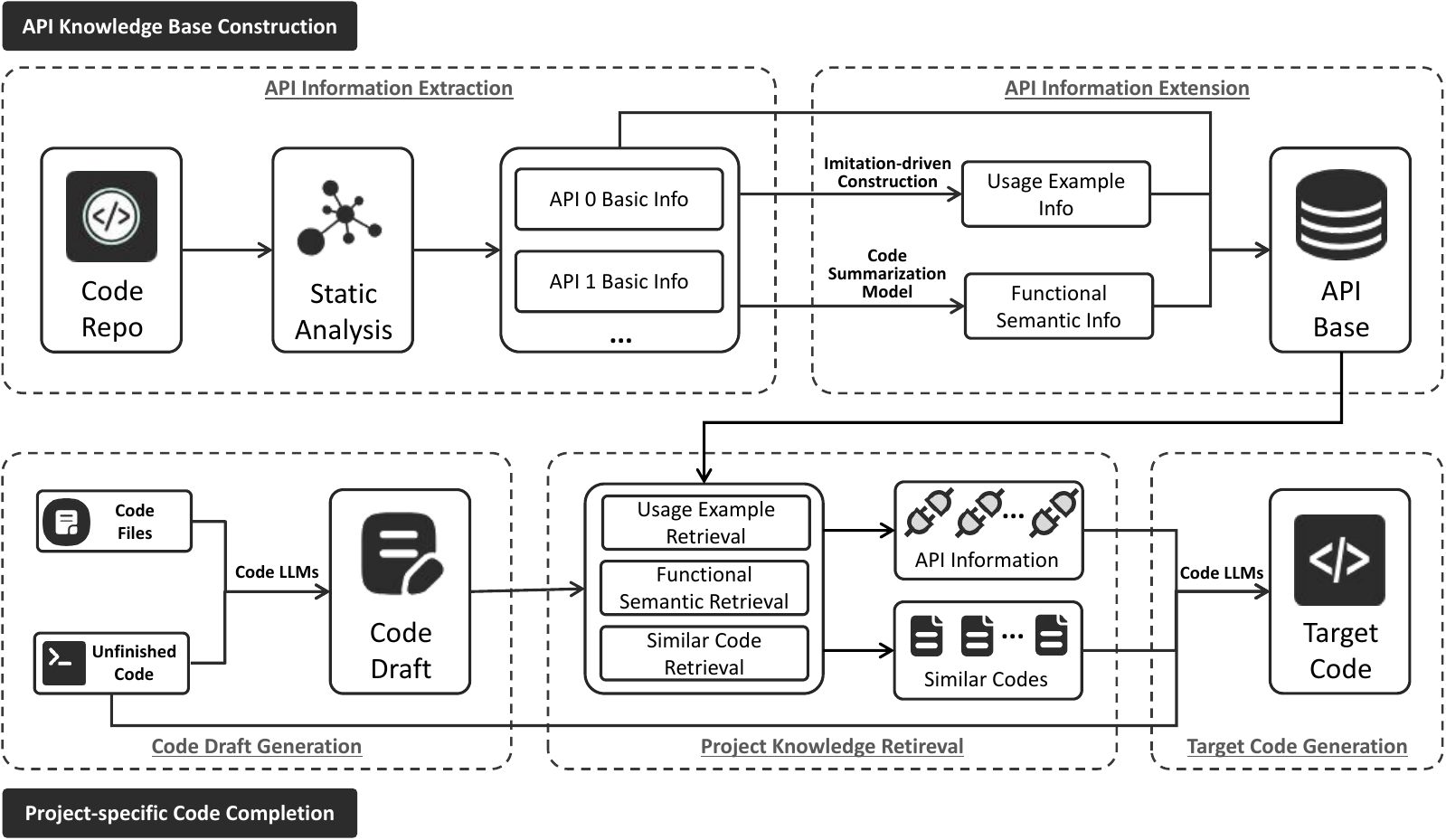}
    \caption{The overall framework of our approach.}
    \label{fig:framework}
\end{figure*}

We focus on project-specific line-level code completion. The framework of our approach is presented in Figure \ref{fig:framework}. 
Our approach takes a source code repository and unfinished code as input. The output of our approach is the target code that continues the unfinished code.
 
In this section, we will detail our approach designed to fully harness the potential knowledge within the project to assist LLMs in code completion through two phases \new{(i.e., API knowledge base construction and project-specific code completion)}. 
In the first phase, we extract the project's APIs and uncover their potential knowledge (i.e., usage examples and functional semantic information) to construct an API database (Section \ref{sec:api_base_instruction}). In the second phase, we describe the three-step code generation pipeline (Section \ref{sec:code_completion}). First, we combine similar code snippets with the unfinished code to generate a code draft (Section \ref{sec:code_draft_generation}). Next, we utilize the code draft to locate the involved internal APIs and similar code snippets (Section \ref{sec:project_knowledge_ret}). Finally, we integrate all retrieved information and the unfinished code into the prompt, which serves as the input for the LLM, ultimately generating target code that fully leverages the project-specific knowledge (Section \ref{sec:target_code_generation}).

\subsection{API Knowledge Base Construction}\label{sec:api_base_instruction}
The upper part of Figure \ref{fig:framework} shows the process of API knowledge base construction.
\new{In this phase, we take a source code repository as input. Through static analysis, we extract the basic information of each API in the repository, including its signature, class, function body, and file path. We then expand the usage example information via imitation-based rewriting and expand the functional semantic information using a code summarization model. Finally, we collect and organize both the basic and extended information for each API to construct the API knowledge base corresponding to the repository.}

\subsubsection{API Information Extraction}
To collect dispersed information within the project, we extract all APIs within the project through static analysis. Specifically, given a code repository, we traverse all the code files within the repository. For each code file, we use tree-sitter~\cite{treesitter} to parse it into an Abstract Syntax Tree (AST) and then identify and extract each function. For each function, we record the following four basic pieces of information: 1) Signature: The function's signature information, including the function name, parameters, and return type. 2) Class: The class to which the function belongs, if applicable. 3) Function Body: The body of the function. 4) File Path: The relative location of the file containing the function within the repository. 

\subsubsection{API Information Extension}
As we analyzed in section \ref{sec:motivation}, simple retrieval of similar snippets cannot accurately locate the specific code of the required API. However, this information is often critical. Fortunately, each API within a code repository contains a wealth of implicit information (i.e., usage example information and functional semantic information) that can be extracted and expanded. This hidden information can assist us in locating the project-specific knowledge needed for code completion. Below, we will demonstrate the specific reasons and methods for expanding these two types of information:

\noindent\textbf{Usage Example Information.} As shown in Figure \ref{fig:motivation_1}, the LLMs call a nonexistent function in the completed code and incorrectly fill in the parameters. However, if we provide the necessary function information to the LLMs, they can correctly call it and fill in the appropriate parameters to complete the correct prediction. The function called in the ground truth is very similar to the initial prediction result of the LLMs, suggesting that we can help the LLMs find the corresponding function through their initial prediction. 
However, differences in function names and parameters
\new{
between the initial prediction and the ground truth
}, as well as the gap between function usage and definition (more information is included in a function definition than the corresponding function usage, such as the def keyword, return type, parameter type, etc.), can hinder our retrieval.

To bridge this gap, we can construct usage examples for each API to imitate real API usages as closely as possible and use an encoder to represent the highly unstructured text as vectors. 
\new{
For each function, we aim to construct code snippets that resemble its typical usage in practice. Due to the flexibility of function invocation, a single function may have multiple usage forms. For example, a static method can be invoked via either \code{ClassName.function()} or \code{class\_name.function()} (the snake\_case naming convention is commonly used in Python~\cite{pep8style}). Additionally, language-specific features introduce further variation in function usage.
For example, Python supports named arguments, allowing developers to optionally specify parameter names during calls.
Exhaustively enumerating all possible invocation patterns for each function will result in significant resource consumption and increased retrieval latency.
To support effective function usage retrieval, we design a small set of heuristic-based rules to generate a few representative usage examples for each function or method. These heuristics aim to approximate the form in which a function is likely to appear during actual usage, considering the variability and syntactic flexibility of different programming languages.
}

\begin{table*}
    \centering
    \caption{Summary of heuristic rules for constructing function/method usage examples. We use \textbf{function} to refer to a callable unit in Python, while in Java, we use \textbf{method} to refer to member functions.}
    \begin{tabular}{lll}
        \toprule[1pt]
         Type & Form & Description \\
         \midrule
         \grayline \multicolumn{3}{c}{Python}\\
         \midrule
         \multirow{4}{*}{Regular Function} & function\_name(arguments) & Unqualified call; direct import usage. \\
         & filename.function\_name(arguments) & Module-qualified usage form. \\
         & function\_name() & Argument-less variant for simplicity. \\
         & filename.function\_name() & Qualified, no-argument form. \\
         \midrule
         \multirow{4}{*}{Class Function} & class\_name.function\_name(arguments) & Instance function call via object. \\
         & ClassName.function\_name(arguments) & Class-level access for static class functions. \\
         & class\_name.function\_name() & No-argument form for object function. \\
         & ClassName.function\_name() & Static class function with no arguments. \\
         \midrule
         \multirow{4}{*}{Constructor} & ClassName(arguments) & Object creation via constructor. \\
         & class\_name = ClassName(arguments) & Assignment upon construction. \\
         & ClassName() & No-argument constructor call. \\
         & class\_name = ClassName() & Assigned form without arguments. \\
         \midrule
         \grayline \multicolumn{3}{c}{Java}\\
         \midrule
         \multirow{3}{*}{Common Method} & className.methodName(arguments) & Instance method call. \\
         & ClassName.methodName(arguments) & Static method call via class. \\
         & TypeName typeName = className.methodName(arguments) & Call preceded by variable declaration for complex types. \\
         \midrule
         \multirow{2}{*}{Constructor} & ClassName className = new ClassName(arguments) & Standard constructor with assignment. \\
         & new ClassName(arguments) & Constructor without assignment. \\
         \midrule
         \multirow{1}{*}{Inner Class} & outerInstance.innerInstance.methodName(arguments) & Invocation of method in nested class context. \\
         \bottomrule[1pt]
    \end{tabular}
    \label{tab:uer_rules}
\end{table*}

\new{
Table~\ref{tab:uer_rules} summarizes the strategies to construct usage examples for both Python and Java. 
For Python, we categorize APIs into three types: regular functions, class functions, and constructors, each with tailored generation strategies. 
For regular functions, we create both unqualified and module-qualified forms, with and without arguments. This design accounts for Python’s flexible import system and invocation patterns while avoiding the combinatorial explosion of generating all possible argument combinations by selecting only two extremes, i.e., all arguments present and none. 
For class functions, we consider whether a method is decorated with \code{@staticmethod} or \code{@classmethod}, and generate both object-based and class-based invocations to capture typical usage in object-oriented contexts. 
For constructors, since object instantiation in Python implicitly invokes \code{\_\_init\_\_}, we generate expressions such as assignment-based and no-assignment-based forms to reflect common instantiation styles.
In Java, a statically typed and object-oriented language, we adopt strategies that align with its stricter syntax and conventional usage patterns. 
For common methods, we generate instance-based and static invocations depending on whether the method can be statically called. 
Additionally, when the return type is complex, we generate variable declarations to mirror realistic usage. 
For constructors, we include both standalone instantiations and assignments, reflecting typical object creation practices in Java. 
Furthermore, recognizing the frequent use of inner classes, we generate forms like \code{outerInstance.innerInstance.methodName (arguments)} to cover nested class usage, which requires explicit instantiation through the enclosing class.
These heuristics are not meant to exhaust all possible call variations, but to capture the most common and informative patterns for retrieval. 
More comprehensive examples are provided in Appendix A~\cite{appendix}.
}

Once we have constructed usage examples (UEs) for all functions, we encode them and save their vector representations for subsequent retrieval steps by using UniXcoder~\cite{guo2022unixcoder}, which is widely used in tasks such as code search.

\begin{figure}[t]
    \includegraphics[width=0.49\textwidth]{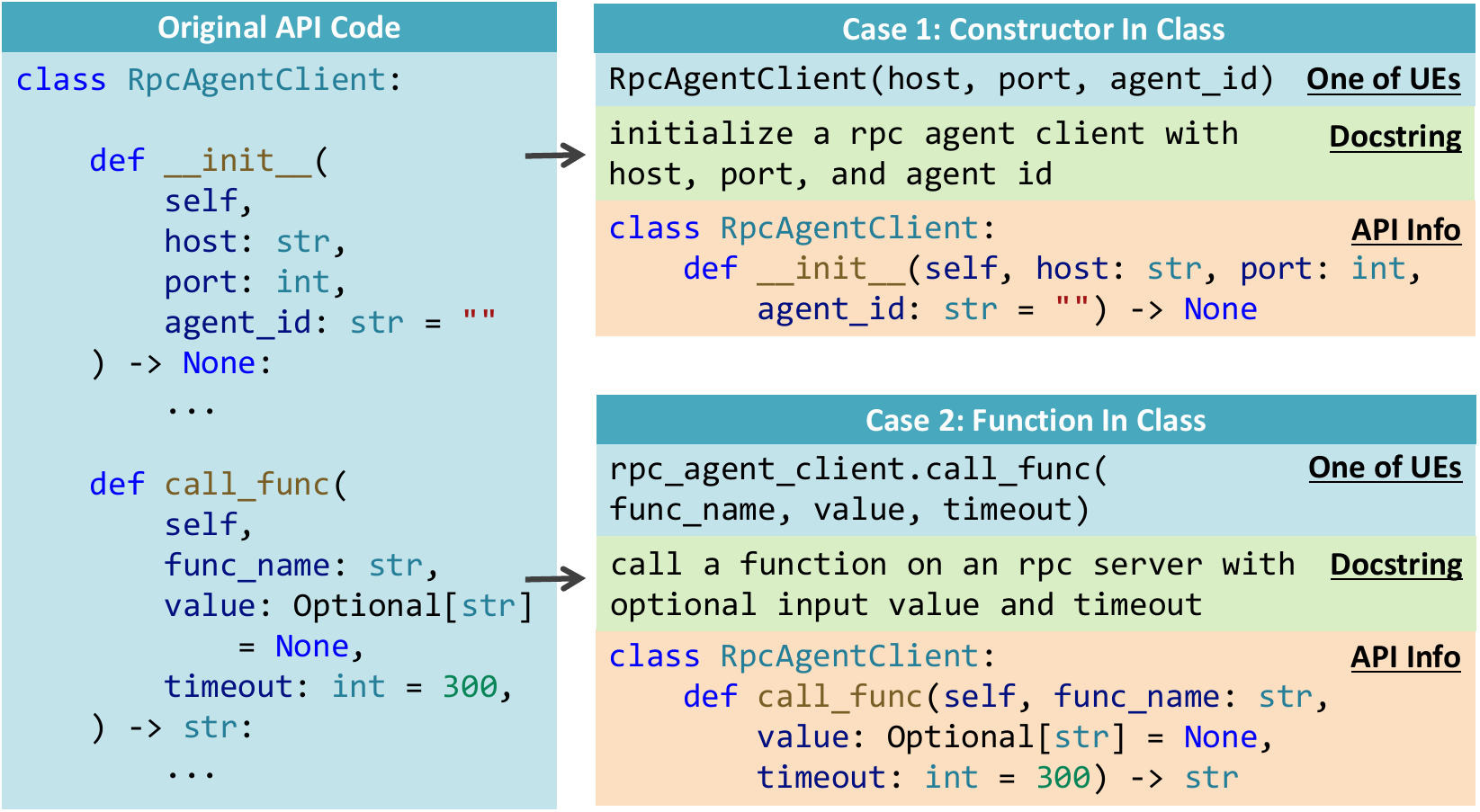}
    \caption{Real API information construction example}
    \label{fig:api_info_example}
\end{figure}

\noindent\textbf{Functional Semantic Information.} As shown in Figure \ref{fig:motivation_2}, the LLMs identify the intention to complete the code but, due to a lack of crucial dependency information, generate a lot of redundant and useless code instead of directly calling the API in the code repository that implements the corresponding functionality. In actual development, since the developer has a certain understanding and knowledge of the project's APIs when they need to implement a functional requirement, they first look for an existing API that can achieve the same functionality to use directly. Two code snippets that implement the same functionality are semantically similar but not always lexically similar. Simple lexical similarity cannot capture the complex semantic information of each API. The APIs in the project are often presented in the form of code without explicitly showing their functional semantic information. Given a code repository, we cannot easily and directly understand the functionality of each API.
 
\begin{figure}[t]
    \includegraphics[width=0.49\textwidth]{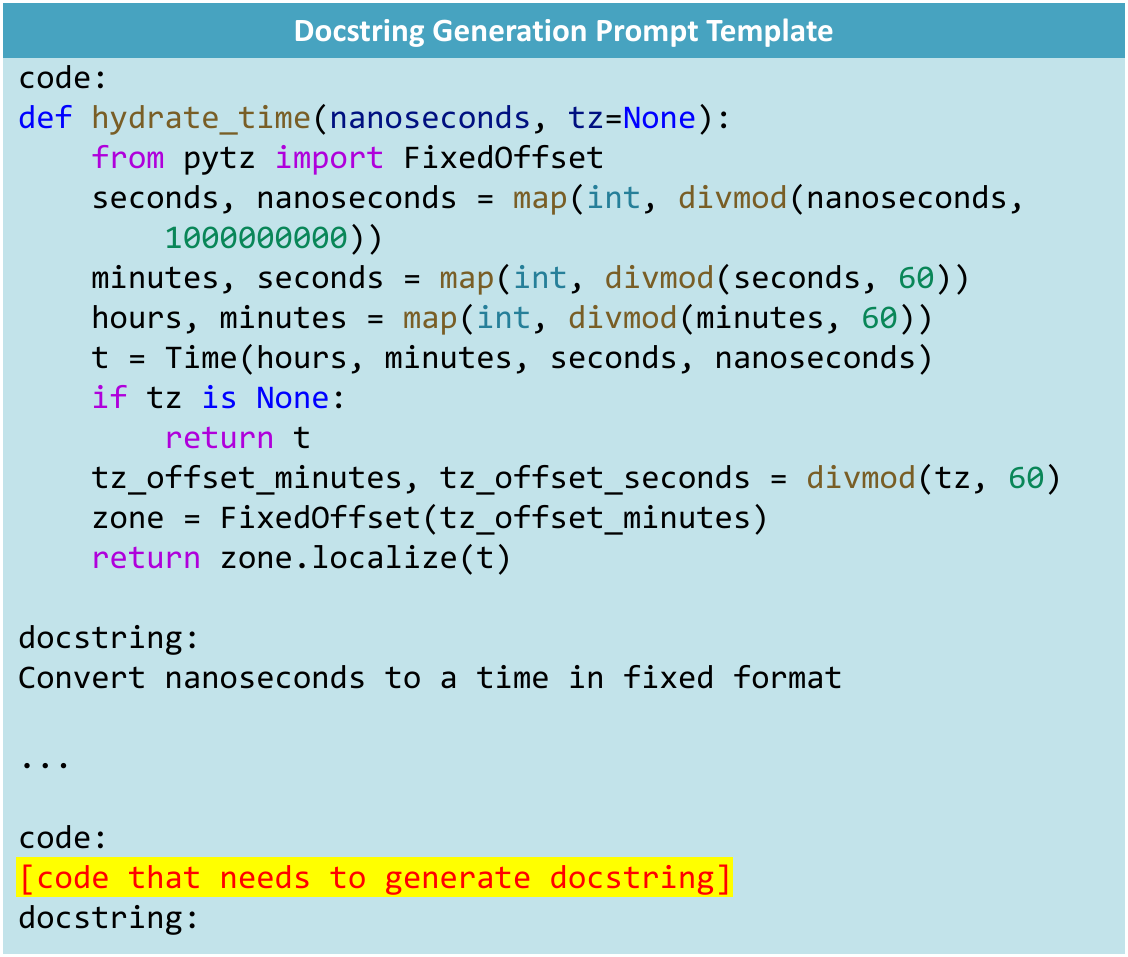}
    \caption{Our summary template, where examples are sourced from CoderEval\cite{yu2024codereval}}
    \label{fig:summary_prompt}
\end{figure}

To address this information gap, we propose to use LLMs to generate a docstring to summarize the function body of each API, which represents the functional semantic information of the API. To ensure that the docstrings generated by the LLMs are of high quality and follow a consistent pattern, we use in-context learning~\cite{dong2022survey,hendel2023context} and include some high-quality code-docstring pairs in the template, which can help the language model understand the task based on the provided examples and produce uniform and high-quality docstrings.
\new{Specifically, we use the prompt template shown in Figure \ref{fig:summary_prompt} to perform code summarization. In our template, we include high-quality code-docstring pairs from CodeEval~\cite{yu2024codereval} as examples (e.g., the code of the function \code{hydrate\_time} along with its natural language docstring) at the top of the prompt, followed by the actual code for which a docstring needs to be generated. This method ensures more uniform output from the model.}
After generating the docstrings, we also use UniXcoder to encode each docstring, saving their vector representations for subsequent retrieval steps.

Eventually, we construct a custom API knowledge base, where each item contains the following fields: signature, class, function body, file path, UE, UE embedding, docstring, and docstring embedding.

\subsection{Project-specific Code Completion}\label{sec:code_completion}
Once the database is constructed, we can proceed with our project-specific code completion pipeline.
\new{The lower part of Figure \ref{fig:framework} illustrates the entire pipeline. In this phase, we take the unfinished code as input and generate a code draft by incorporating similar code snippets from other files. This draft is then used to guide usage examples retrieval and functional semantic retrieval, helping to infer potential API information needed for completion. Meanwhile, we also retrieve code that is similar to the code draft. Finally, we organize all the gathered information to construct a new prompt for target code generation.}

\subsubsection{Code Draft Generation}\label{sec:code_draft_generation}
The code draft is the foundation of the three-stage code completion method, providing an initial solution for subsequent retrieval and generation stages to reference and improve upon. Understanding the in-file context is crucial during the code generation process~\cite{clement2021long}. In-file information, such as existing function definitions, variable names, and class structures, provides the context of the current file, enabling the model to generate code that better fits the current coding environment. Additionally, previous works~\cite{ding2024crosscodeeval,zhang2023repocoder,wu2024repoformer,li2024deveval} have demonstrated that LLMs can simply mimic similar code snippets, bringing the prediction results closer to the ground truth. The code draft does not aim for accurate predictions but rather seeks to offer an initial reference and improvement solution for subsequent steps. Therefore, in this step, we simply use both in-file information and the code in the code files within the repository similar to the unfinished code to construct the prompt for generating our code draft. Specifically, following RepoCoder~\cite{zhang2023repocoder}, we allocate half of the prompt length to the retrieved snippets and the other half to the in-file information. We retrieve a set of similar code snippets from a repository via a fixed-size sliding window and then obtain the code snippet in the subsequent window. This method is also mentioned in RepoCoder~\cite{zhang2023repocoder}. The Jaccard index~\cite{jaccard1912distribution} is used to measure the similarity between two code snippets. The definition is as follows:

$$
JaccardIndex(A,B)=\frac{| A \cap B |}{| A \cup B |}
$$
where \( A \) and \( B \) represent two sets of tokens from the respective code snippets. The Jaccard index measures the similarity between two code snippets based on the common tokens they share.

Notably, in code completion, it is common practice to have the model predict several tokens and then perform post-processing on the prediction results to obtain the final output based on the desired code completion level (i.e., line level and function level)~\cite{liu2023repobench,ding2024crosscodeeval,yu2024codereval}. In our code draft, we do not perform post-processing but retain the model's original output.

\subsubsection{Project Knowledge Retrieval}\label{sec:project_knowledge_ret}
As described in Section \ref{sec:motivation}, due to the lack of awareness of project knowledge within the project, the predictions of LLMs often deviate from the actual requirements (i.e., generating incorrect dependencies and redundant code). To bridge the gap between the code draft and the ground truth, we obtain the necessary top-k API information based on usage example retrieval (UER) and functional semantic retrieval (FSR).

\noindent\textbf{Usage Example Retrieval.} To address the issue of incorrect dependency usage in the code draft, we search for APIs that are similar in identifier names. Specifically, we extract the line in the code draft that needs to be completed (e.g., the first line of code in the red box \new{labeled \textit{Generation Result}} in Figure \ref{fig:motivation_1}) and use an embedding model to generate its vector representation as the query. We then calculate the cosine similarity~\cite{salton1988term} between the vector representations of the query and each API's UE. Finally, we select the top-k APIs with the highest scores.

\noindent\textbf{Functional Semantic Retrieval.} To solve the problem of not utilizing existing APIs in the code draft, we search for APIs that are functionally similar to the code draft. Specifically, we extract the code from the point needing completion to the end from the code draft (e.g., all the code in the red box \new{labeled \textit{Generation Result}} in Figure \ref{fig:motivation_2}), process it into a docstring (as described in section \ref{sec:api_base_instruction}), encode the obtained code docstring, and calculate the cosine similarity with the docstring of each API. Finally, we select the top-k APIs with the highest scores.

\noindent\textbf{Similar Code Retrieval.} Similar snippets within the same project are valuable for code completion, even if they are not entirely replicable. In this step, we also retrieve similar code snippets. Following RepoCoder, we no longer use the unfinished code as the query but instead use the code draft, because the code draft is closer to the ground truth compared to the unfinished code. We use the Jaccard index to calculate the similarity between the code draft and the candidate code snippets. Then, we obtain a list sorted by scores. Due to the potentially large differences in length between code snippets, we no longer use the top-k method. Instead, we get code snippets from the highest to the lowest scores until the preset context length is filled.

\subsubsection{Target Code Generation}\label{sec:target_code_generation}
\begin{figure}[t]
    \includegraphics[width=0.49\textwidth]{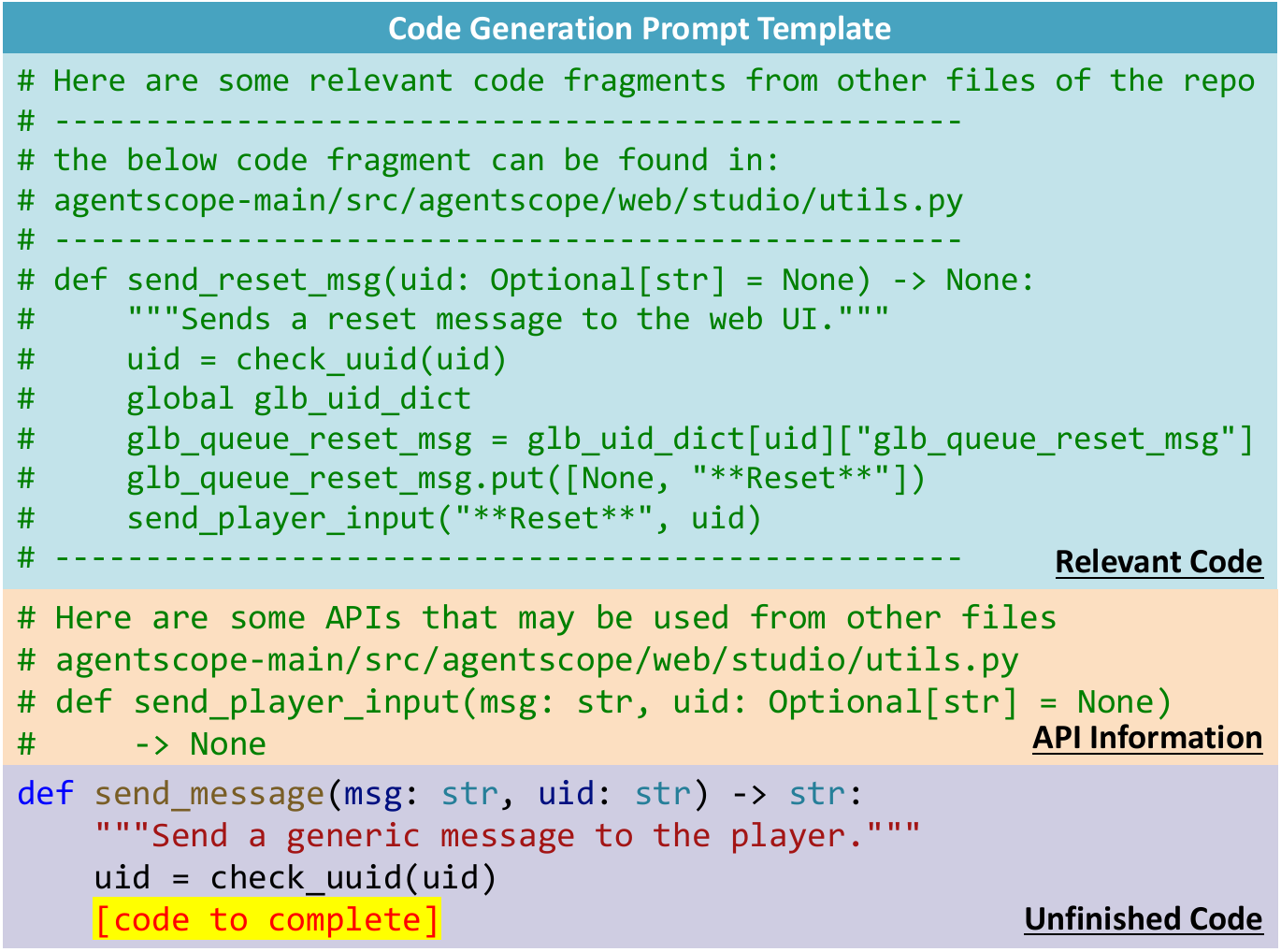}
    \caption{A visual example of our approach prompt format}
    \label{fig:generation_prompt}
\end{figure}

Finally, we describe how to utilize the results from the previous steps and invoke LLMs to complete code generation. For each retrieved internal API, we collect its definition and file path, and extract key elements such as the function name, parameter list, and return type. 
\new{
These elements serve as high-quality contextual cues to inform the model about expected usage patterns. It is important to note that our goal is not to ensure the correctness of specific argument values, but rather to provide sufficient semantic context to guide the model’s generation. The actual instantiation of parameter values is deferred to the language model.
}
\new{
As shown in the orange boxes labeled \textit{API Info} in Figure~\ref{fig:api_info_example}, we combine relevant API details to simulate the definition context. Specifically, we concatenate the function signature with its enclosing class definition.
}
For the results obtained from similar code retrieval, we similarly obtain the corresponding code and file path for each snippet. Figure \ref{fig:generation_prompt} illustrates our prompt design.
\new{Given a task to complete the rest body of the \code{send\_message} function, we obtain a code snippet from a similar function, namely the source code of \code{send\_reset\_msg}, by performing similar code retrieval. Through internal API retrieval, we also obtain an API that may be useful for the completion, \code{send\_player\_input}. We construct the prompt for target code generation by first placing the file path and source code of \code{send\_reset\_msg}, followed by the file path and API information of \code{send\_player\_input}, and finally appending the unfinished code. This combined context serves as the prompt for code generation.}

\section{Experimental Setup}
\subsection{Dataset Construction}\label{sec:dataset_construction}
Our task is real-world project-specific line-level code completion, which should involve cross-file information and internal dependencies. Although some code completion benchmarks utilize project-level information~\cite{liu2023repobench,shrivastava2023repository,shrivastava2023repofusion,ding2022cocomic,ding2024crosscodeeval, zhang2023repocoder}, they have the following issues. 1) Data leakage: some benchmarks contain some code that is also used to train LLMs, possibly affecting the reliability of the evaluation~\cite{shrivastava2023repofusion,shrivastava2023repository,ding2022cocomic,zhang2023repocoder}. 2) Lack of representativeness: the selected projects of some benchmarks are not popular enough and may have low code quality or follow non-standard practices~\cite{ding2024crosscodeeval,liu2023repobench}.
3) Misalignment with typical usage: before an internal API is used for the first time within a project, its corresponding import statement may not exist. Previous works~\cite{liu2023repobench,shrivastava2023repository,shrivastava2023repofusion,ding2022cocomic,ding2024crosscodeeval, zhang2023repocoder} have not considered that the leakage of import statements can cause evaluation results to deviate from actual applications~\cite{semenkin2024full}. To avoid these potential risks, we construct a new benchmark named ProjBench. It includes Python and Java, the two most popular languages. For each language, ProjBench contains 10 real-world projects.

\subsubsection{Project Selection}
\new{
To avoid potential data leakage, we select projects based on their creation time rather than the latest release dates, as some projects may evolve slowly and retain stable APIs.
}
Specifically, we choose non-forked repositories on GitHub that are created after January 1, 2024. The number of stars indicates the popularity of the repositories, and more popular repositories often have high attention and active maintenance, ensuring a certain level of project quality. So we only collect the code repositories with more than 1000 stars for Python and 100 stars for Java. We use a different star limit for Java because we are unable to obtain a sufficient number of projects. To ensure that our benchmark covers different domains, we manually identify whether each project is a fork or copy from another project and only choose one from the fork and source projects to retain. For example, MagicClothing is a branch version of OOTDiffusion, so we retain only one of them. 
\new{
Finally, inspired by the design of RepoCoder~\cite{zhang2023repocoder}, we randomly select 10 projects each for Python and Java to balance diversity and evaluation efficiency. 
}
More information about the selected repositories can be found in Table \ref{tab:repo_benchmark}.
\begin{table}[h]
    \centering
    \caption{The projects used for our benchmark.}
    \begin{tabular}{l c c c}
        \toprule[1pt]
         Project & Created & Files & Lines\\
         \midrule
         \grayline \multicolumn{4}{c}{Python}\\
         agentscope & 2024-01-12 & 157 & 21939\\
         arida & 2024-01-10 & 113 & 2306 \\
         corenet & 2024-04-18 & 490 & 60337\\
         GPT-SoVITS & 2024-01-14 & 110 & 20045\\
         logfire & 2024-04-23 & 124 & 23239\\
         MagicClothing & 2024-02-06 & 282 & 43690\\
         open-parse & 2024-03-22 & 57 & 8503\\
         penzai & 2024-04-04 & 126 & 32924\\
         QAnything & 2024-01-03 & 172 & 23894\\
         skyvern & 2024-02-28 & 104 & 12584\\
        \midrule
        \grayline \multicolumn{4}{c}{Java}\\
        ddd-boot & 2024-01-15 & 419 & 29355\\
        aShellYou & 2024-01-23 & 46 & 7497\\
        kspider & 2024-07-07 & 168 & 15105\\
        WaEnhancer & 2024-04-30 & 103 & 13285\\
        kafka-ui & 2024-01-22 & 410 & 34529\\
        netty-mqtt-client & 2024-05-13 & 82 & 10706\\
        warm-flow & 2024-01-02 & 250 & 26659\\
        CatVodSpider & 2024-03-10 & 141 & 14536\\
        online-exam-system & 2024-06-01 & 208 & 13708\\
        Tubular & 2024-01-21 & 376 & 60175\\
         \bottomrule[1pt]
    \end{tabular}
    \label{tab:repo_benchmark}
\end{table}

\subsubsection{Benchmark Construction}
To ensure that our samples involve project-specific knowledge during completion, we perform static analysis on the entire project to find all code lines involving cross-file information and internal dependencies. 
\new{
To ensure that the APIs we identify and use in our method are truly project-specific, we identify internal imports from third-party or public libraries during dataset construction through static analysis.
}
Specifically, we used tree-sitter to parse each code file, identify all import statements related to custom modules and packages, and further collect code lines that used these modules and packages within the file. To closely mimic real development scenarios, for samples where the code line is the first use of cross-file dependencies in the current file, we identify the corresponding import statements and mask them. 
\new{
This masking prevents models from relying on explicit API declarations and forces them to infer internal API usage from contextual clues, making the completion task both more realistic and more challenging.
}
Following CrossCodeEval~\cite{ding2024crosscodeeval}, we exclude examples if the references are found verbatim in any other source code file within the repository. Finally, we randomly selected 200 lines that involved project-specific knowledge from each repository. 
 
ProjBench focuses on the task of line-level code completion. In real IDEs, we often face two line-level completion scenarios: 1) single-line~\cite{shrivastava2023repository,shrivastava2023repofusion,ding2022cocomic,ding2024crosscodeeval}, i.e., the developer has written a few tokens and need the model to complete the remaining tokens until the end of the line; 2) next-line~\cite{liu2023repobench,zhang2023repocoder}, i.e., the developer has finished writing one line of code and need the model to complete the next line. To cover both scenarios, we randomly select a tree-sitter token from the sample that appears before a cross-file token as the cursor position, allowing the model to complete from the cursor position to the end of the line.

Finally, we construct a benchmark that aligns with real-world scenarios. It includes projects with large-scale contexts, averaging 197 code files and approximately 23751 lines of code per sample.

\vspace{1em}

\new{
Besides using a newly constructed benchmark specifically designed to evaluate our approach, we also incorporate an existing and widely-used~\cite{liang2024repofuse,wang2024rlcoder} benchmark, CrossCodeEval (CCEval)~\cite{ding2024crosscodeeval}, to further validate the generality and robustness of our method. Incorporating CCEval enables us to compare our approach against prior work on a well-established dataset and to demonstrate its effectiveness across diverse evaluation scenarios.
}
    
\subsection{Baseline Methods}
Our approach is designed for seamless integration with existing LLMs, necessitating only black-box access to these models. To evaluate the effectiveness of our approach, we conduct comparative analyses against the following baselines:

\begin{itemize}
    \item \textbf{InFile:} Utilizing in-file context is highly valuable for code completion scenarios\cite{clement2021long}. This method leverages the code within the current file to fill the prompt up to the maximum context length and does not provide additional context. Then, the prompt is directly fed into the code LLM to obtain the completion results.
    \item \textbf{RepoFuse~\cite{liang2024repofuse}:} 
    \new{
    This method enhances code completion by integrating two types of context: rationale context and analogy context. The rationale context analysis component analyzes the import relationships within the file containing the unfinished code, extracting dependencies. The analogy context retrieval component leverages BM25~\cite{robertson2009probabilistic} to retrieve semantically similar code snippets from other files. All candidate contexts, both rationale-based and analogy-based, are ranked using UniXcoder embeddings and cosine similarity. The top-ranked contexts are then concatenated with the unfinished code as the final input to the model in order.
    }
    \item \textbf{RepoCoder~\cite{zhang2023repocoder}:} It is a state-of-the-art framework for project-specific code generation. The baseline employs an iterative retrieval and generation approach to generate the target code. This method involves retrieving similar code snippets based on the previously generated outputs. In each iteration, RepoCoder concatenates the similar code snippets retrieved in the last step with the unfinished code and calls the code LLM to generate the result.
\end{itemize}

\subsection{Implementation Details}\label{sec:implementation_details}
In this study, we utilize several mainstream code language models (i.e. DeepSeekCoder-6.7B~\cite{guo2024deepseek}, CodeLlama-7B~\cite{roziere2023code}, and StarCoder2-7B~\cite{lozhkov2024starcoder}) as our base models.
All models involved in the experiments are open-source, and we obtain them via Hugging Face's Transformers library~\cite{wolf2020transformers} and use them via vLLM~\cite{kwon2023efficient}.
For all code generation processes, we set the model's maximum context length to 4096 and use a maximum generation length of 128. The length of the retrieved code snippets is set to occupy half of the prompt length. For project dependency retrieval, we set k to 4, which means we will obtain information for a total of 8 APIs. For summarizing code, we use Llama3-Instruct-8b.
\new{
We implement the InFile baseline by ourselves due to its simplicity. For all other baselines, we use their official implementations released by the original authors on GitHub.
}
For the sliding window methods involved in the experiments, we follow RepoCoder~\cite{zhang2023repocoder}, setting the window length to 20 lines and the sliding size to 10.

\new{
It is worth noting that, given RepoFuse's reliance on import statements, we do not mask such information in both ProjBench and CCEval to avoid hindering its performance. In contrast, our approach is evaluated under more realistic and challenging conditions, 
i.e., without relying on import statements as retrieval hints and completion hints. This better reflects real-world development scenarios where such information may be incomplete or unavailable. 
Although this setting is unfair for our method, it ensures that RepoFuse is correctly implemented and can further highlight the effectiveness of our approach.
}

\subsection{Evaluation Metrics}

Following the previous work~\cite{zhang2023repocoder,ding2024crosscodeeval,ding2022cocomic,liang2024repofuse}, we evaluate the performance of our approach and the baselines on two dimensions: code match and identifier match (ID match).

\noindent\textbf{Code Match:} Code match directly uses Exact Match (EM) and Edit Similarity (ES)~\cite{levenshtein1966binary} to compare the generated code with the ground truth. EM is a binary metric used to evaluate how accurately a model's prediction exactly matches the ground truth. ES evaluates how similar two strings are by calculating the minimum number of edit operations required to transform one string into another.

\noindent\textbf{ID Match:} This dimension can help evaluate whether the model correctly applies project-specific knowledge. Specifically, we extract the identifiers in the model predictions and ground truth by constructing the code's Abstract Syntax Tree (AST) and then compare them to obtain the EM and F1 scores at the identifier level.

\section{Evaluation Results}
In this section, we report and analyze the experimental results to answer the following research questions (RQs):
\begin{itemize}
    \item \textbf{RQ1: How does our approach perform compared with other methods for project-specific code completion tasks?} We compare our approach with baselines including Infile, RepoFuse, and RepoCoder to verify the effectiveness of our approach.
    \item \textbf{RQ2: How do the two components of our approach (i.e., UER and FSR) respectively contribute to overall effectiveness?} We start from a minimal baseline without UER and FSR, then incrementally add each component to observe the relative improvements in performance.
    \item \textbf{RQ3: How useful is it to integrate our UER and FSR into existing methods?} We incorporate UER and FSR into the baseline and observe the performance changes of each baseline.
\end{itemize}
\begin{table*}[ht]
    \centering
    \caption{Evaluation results of different methods}
    \begin{tabular}{cclcccccccccccc}
        \toprule[1pt]
         \multirow{3}{*}{Benchmark} & \multirow{3}{*}{Lang} & \multirow{3}{*}{Method} & \multicolumn{4}{c}{DeepSeekCoder-6.7B} & \multicolumn{4}{c}{CodeLlama-7B} & \multicolumn{4}{c}{StarCoder-7B}\\
         \cmidrule(lr){4-7}\cmidrule(lr){8-11}\cmidrule(lr){12-15}
                                                    & & & \multicolumn{2}{c}{Code Match} & \multicolumn{2}{c}{ID Match} & \multicolumn{2}{c}{Code Match} & \multicolumn{2}{c}{ID Match} & \multicolumn{2}{c}{Code Match} & \multicolumn{2}{c}{ID Match}\\
                                                    & & & EM & ES & EM & F1 & EM & ES & EM & F1 & EM & ES & EM & F1\\
         \midrule
         \multirow{8}{*}{Projbench} & \multirow{4}{*}{Python} 
         & InFile & 18.45 & 45.39 & 23.90 & 41.58 & 21.20 & 48.47 & 26.55 & 44.23 & 18.80 & 44.68 & 24.05 & 41.57\\
         & & RepoFuse & 23.90 & 49.47 & 30.15 & 47.03 & 29.80 & 55.83 & 35.70 & 53.62 & 25.90 & 51.65 & 31.70 & 48.94\\
         & & RepoCoder & 24.45 & 50.01 & 30.75 & 48.06 & 28.95 & 54.77 & 34.75 & 52.44 & 25.15 & 51.06 & 31.20 & 49.05\\
         & & Ours & \textbf{29.79} & \textbf{53.19} & \textbf{36.10} & \textbf{51.76} & \textbf{36.76} & \textbf{59.81} & \textbf{42.86} & \textbf{58.83} & \textbf{31.55} & \textbf{55.87} & \textbf{37.71} & \textbf{53.39} \\
         \cmidrule(lr){2-15}
         & \multirow{4}{*}{Java} 
         & InFile & 22.55 & 52.96 & 25.45 & 49.71 & 23.75 & 54.61 & 26.50 & 50.92 & 22.90 & 54.44 & 25.80 & 51.41\\
         & & RepoFuse & 27.00 & 54.98 & 31.35 & 52.25 & 29.35 & 58.86 & 33.05 & 56.16 & 27.20 & 55.65 & 31.10 & 52.41\\
         & & RepoCoder  & 29.00 & 57.20 & 33.30 & 55.78 & 31.00 & 59.20 & 34.70 & 57.11 & 29.65 & 58.45 & 34.25 & 56.37\\
         & & Ours       & \textbf{33.15} & \textbf{58.75} & \textbf{37.70} & \textbf{58.26} & \textbf{34.30} & \textbf{61.40} & \textbf{38.40} & \textbf{59.96} & \textbf{32.40} & \textbf{60.02} & \textbf{36.95} & \textbf{58.52}\\
         \midrule

         \multirow{8}{*}{CCEval} & \multirow{4}{*}{Python} 
            & InFile & 9.08 & 51.32 & 15.87 & 48.01 & 7.02 & 49.55 & 14.11 & 45.39 & 7.32 & 49.96 & 14.18 & 46.33 \\
            & & RepoFuse & 26.79 & 64.11 & 37.00 & 63.52 & 23.56 & 61.18 & 33.36 & 60.28 & 24.32 & 62.53 & 34.22 & 61.64 \\
            & & RepoCoder & 26.87 & 64.55 & 37.60 & 64.39 & 23.64 & 61.91 & 33.58 & 61.38 & 24.54 & 63.10 & 34.86 & 62.52 \\
            & & Ours & \textbf{35.44} & \textbf{69.43} & \textbf{46.36} & \textbf{70.45} & \textbf{32.09} & \textbf{66.71} & \textbf{42.53} & \textbf{67.61} & \textbf{33.00} & \textbf{68.09} & \textbf{43.92} & \textbf{68.63} \\
         \cmidrule(lr){2-15}
         & \multirow{4}{*}{Java} 
            & InFile & 10.66 & 52.90 & 17.72 & 51.32 & 9.44 & 53.78 & 17.06 & 51.75 & 10.14 & 54.70 & 18.23 & 52.72 \\
            & & RepoFuse & 22.95 & 57.76 & 31.18 & 57.27 & 21.93 & 59.97 & 30.72 & 59.35 & 21.51 & 59.43 & 30.58 & 58.27 \\
            & & RepoCoder & 25.85 & 61.02 & 35.25 & 61.80 & 24.40 & 60.86 & 33.24 & 60.88 & 24.82 & 62.38 & 34.50 & 62.46 \\
            & & Ours & \textbf{31.23} & \textbf{63.24} & \textbf{41.23} & \textbf{64.59} & \textbf{31.23} & \textbf{64.38} & \textbf{40.86} & \textbf{65.33} & \textbf{29.92} & \textbf{65.05} & \textbf{39.93} & \textbf{65.59} \\
         \bottomrule[1pt]
    \end{tabular}
    \label{tab:effectiveness}
\end{table*}

\subsection{Effectiveness of Our Approach (RQ1)}
To evaluate the effectiveness of our approach, we compare it with the baselines, i.e., InFile, RepoFuse, and RepoCoder. The comparison results are shown in Table \ref{tab:effectiveness}.
 
According to the results in Table \ref{tab:effectiveness}, we can observe that the InFile baseline performs worse than the other baselines. Because the in-file context fails to provide crucial cross-file context information, highlighting the importance of project-specific knowledge. Across all models, our approach outperforms other RAG-based methods on both Java and Python completion tasks. 
\new{Specifically, compared to the best-performing baseline, our approach achieves an average improvement of +5.91 points in code match EM and +6.26 points in ID match EM.}
\new{
Even though RepoFuse incorporates contextual information by analyzing the import relationships of the file to be completed, which may cause information leakage, its performance remains suboptimal. 
This can be attributed to three main reasons. 
First, its analogy context retrieval is conducted in a single-round manner. Compared to methods such as RepoCoder and ours, it is less effective in retrieving the most semantically similar code snippets. 
Second, its rational context analysis component tends to extract an excessive amount of dependency-related information. While RepoFuse introduces a re-ranking step to mitigate the impact, it still introduces substantial noise unrelated to the completion task. For example, when retrieving information from a dependent class, it indiscriminately includes all defined attributes and methods, many of which are irrelevant. 
Third, we also observe that RepoFuse performs worse on Java compared to Python due to limitations in its rational context analysis. While it uses Jedi for Python, which captures rich semantic relations (e.g., Calls, Overrides), its Java analysis relies on Tree-sitter, which only supports basic Import relations. This restricts its ability to retrieve meaningful contextual information in Java projects, leading to reduced performance.
Compared to RepoCoder, our approach consistently outperforms it across datasets and metrics by explicitly inferring internal API information through UER and FSR. 
In addition to retrieving globally similar code based on surface patterns as done by RepoCoder, our approach further targets internal API semantics, enabling more accurate and context-aware completions.
}

These results indicate that our method is capable of retrieving more comprehensive and precise project-specific knowledge, alleviating the limitations of relying solely on similarity retrieval strategies. It is worth noting that the improvements of our method in ES and F1 scores are less than those in EM. This is because the benefit of our approach lies in helping the LLM accurately generate internal API calls, which are only part of the code snippet to be completed. EM considers the code snippet as a whole and gives binary results, while ES and F1 take into account partial matches between the generated result and the reference. Thus, the gains in ES and F1 are relatively modest.

\begin{figure}
    \centering
    \includegraphics[width=0.49\textwidth]{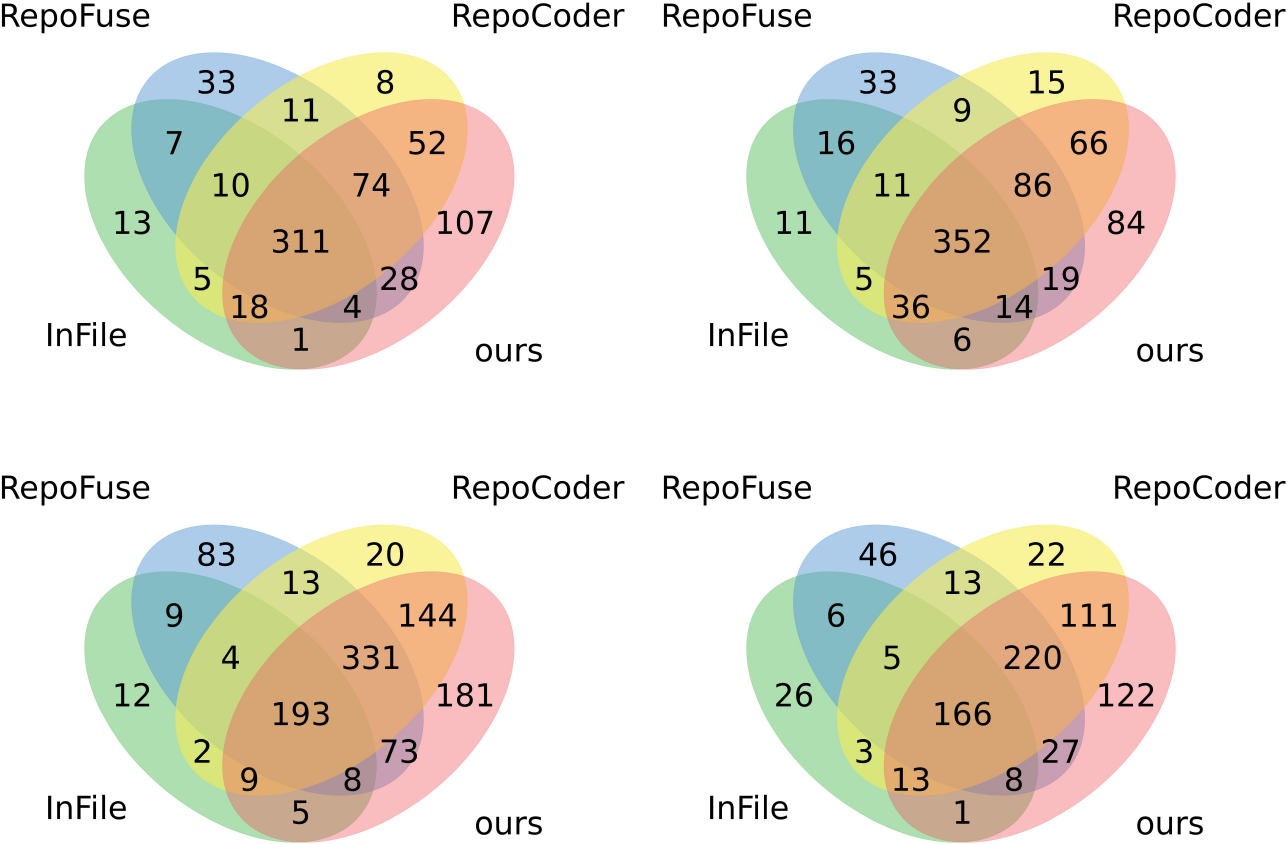}
    \caption{Venn diagrams of EM results for Python (left) and Java (right) on DeepSeekCoder-6.7B. The upper two diagrams correspond to ProjBench, while the lower two correspond to CCEval.}   
    \label{fig:venn}
\end{figure}

\begin{figure*}
    \centering
    \includegraphics[width=0.95\textwidth]{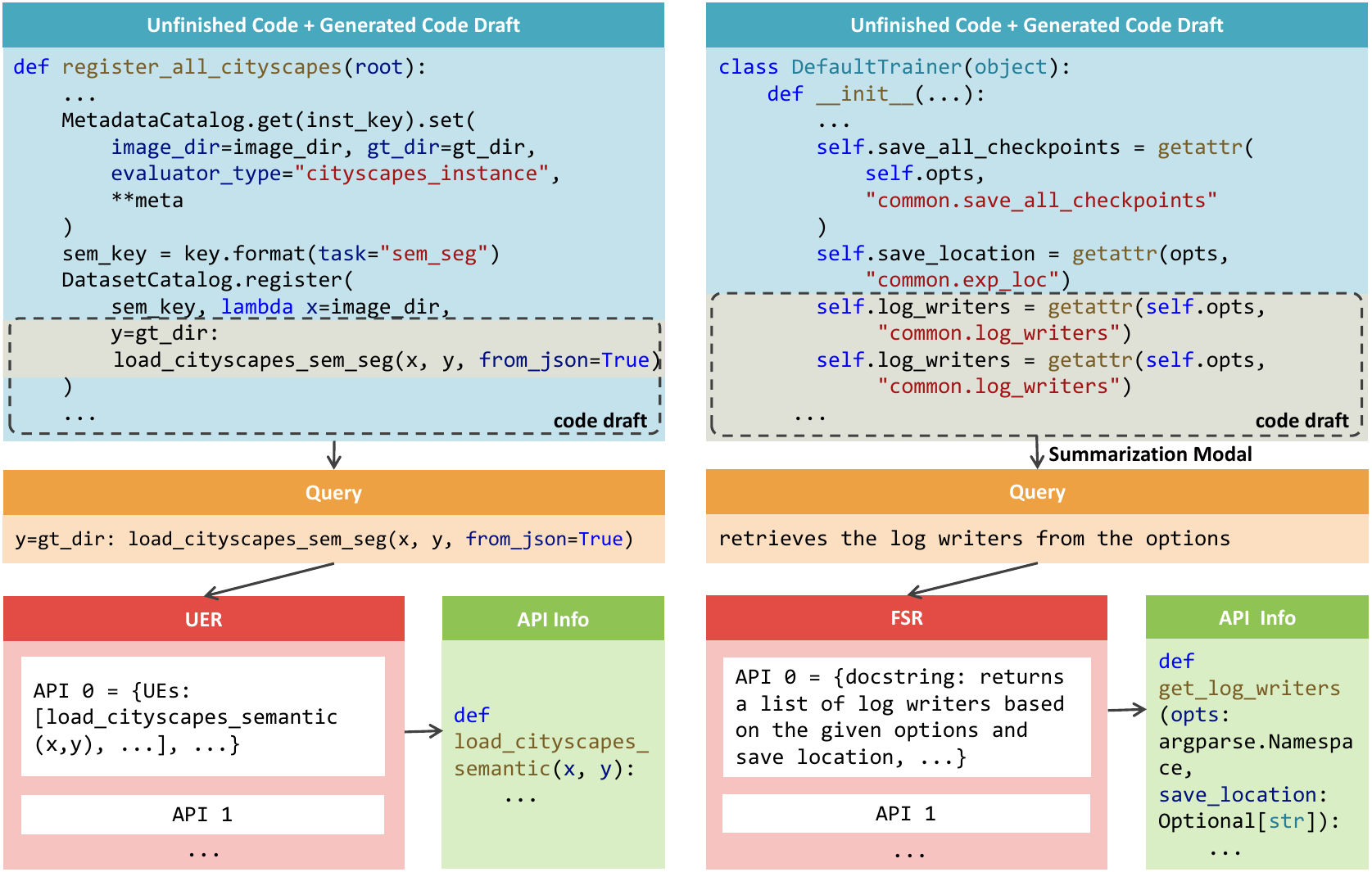}
    \caption{Two test samples, the left from MagicClothing and the right from corenet}
    \label{fig:case_study}
\end{figure*}

To further investigate the differences between the methods, we draw Venn diagrams to show the number of unique samples that are correctly completed by each method, as illustrated in Figure \ref{fig:venn}. Our approach completes the highest number of unique samples on both Projbench and CCEval. Upon manually inspecting the experimental results, we find that the superiority of our method mainly stems from its ability to retrieve the API information relevant to the completion samples. As an example, Figure~\ref{fig:case_study} shows how our approach handles the motivating examples mentioned in Section~\ref{sec:motivation}.

For the first motivating example (the left part of Figure \ref{fig:case_study}), the UER component of our approach successfully retrieves the \code{load\_cituscapes\_sem\_seg} function in the repository based on the similarity between the UE of this function and the function appearing in the code draft (\code{y=gt\_dir: load\_cituscapes\_sem\_seg(x, y, from\_json=True)}). 
When we provide the definition of the retrieved function to the LLM, the LLM correctly completes the task.
For the second motivating example (the right part of Figure \ref{fig:case_study}), the FSR component of our approach successfully retrieves the \code{get\_log\_writers} function in the repository based on the similarity between the FS of this function and the docstring of the code draft (\code{retrieves the log writers from the options}). After providing the definition of this function to the LLM, the LLM completes the task. 
 
Our novel project knowledge retrieval method retrieves the necessary API information for completion without relying on import statements, and by mining the latent representations of the APIs, it outperformed methods like RepoCoder, which rely on superficial similarities between code snippets.

\find{
{\bf RQ1:}
In summary, our approach outperforms the three baselines, exceeding the best-performing baseline by +5.91 and +6.26 in terms of exact code match and exact identifier match, respectively.
}

\subsection{Contributions of Each Component (RQ2)}
\begin{table}
    \centering
    \setlength{\tabcolsep}{4pt}
    \caption{Evaluation results of different variants of our approach on DeepSeekCoder-6.7B.}
    \begin{tabular}{cclcccc}
        \toprule[1pt]
        \multirow{2}{*}{Benchmark} & \multirow{2}{*}{Lang} & \multirow{2}{*}{Variant} & \multicolumn{2}{c}{Code Match} & \multicolumn{2}{c}{ID Match}\\
                                & & & EM & ES & EM & F1\\
         \midrule
         \multirow{6}{*}{ProjBench} & \multirow{3}{*}{Python} & Base & 24.30 & 50.28 & 30.90 & 48.08\\
         & & +FSR & 27.34 & 51.80 & 33.65 & 50.15\\
         & & +UER & 29.25 & 52.76 & 35.45 & 51.18\\
         \cmidrule(lr){2-7}
          & \multirow{3}{*}{Java} & Base & 28.95 & 56.72 & 33.50 & 55.49\\
          & & +FSR & 30.60 & 57.73 & 35.10 & 56.81\\
          & & +UER & 32.85 & 58.63 & 37.45 & 57.90\\
         \midrule
         \multirow{6}{*}{CCEval} & \multirow{3}{*}{Python} & Base & 26.53 & 64.39 & 37.30 & 64.26\\
         & & +FSR & 29.62 & 65.96 & 40.32 & 66.46\\
         & & +UER & 34.67 & 68.89 & 45.63 & 69.92\\
         \cmidrule(lr){2-7}
          & \multirow{3}{*}{Java} & Base & 25.62 & 60.47 & 34.32 & 61.15\\
          & & +FSR & 27.49 & 61.42 & 36.51 & 62.31\\
          & & +UER & 31.23 & 62.99 & 41.00 & 64.71\\
         \bottomrule[1pt]
    \end{tabular}
    \label{tab:ablation}
\end{table}

\new{
To better understand how each component contributes to performance, we conduct an incremental study starting from a minimal baseline: 1) \textbf{Base}: A minimal version without usage example retrieval (UER) and functional semantic retrieval (FSR). 2) \textbf{+UER}: Base enhanced with the usage example retrieval component. 3) \textbf{+FSR}: Base enhanced with the functional semantic retrieval component. This design allows us to isolate the individual impact of each module on performance improvement.
}

The experimental results are shown in Table \ref{tab:ablation}. 
Adding either component significantly improves performance in both code matching and ID matching metrics. 
\new{Specifically, incorporating the usage example retrieval (UER) component leads to an average increase of +5.65 in code match EM and +5.88 in ID match EM. 
Adding the functional semantic retrieval (FSR) component results in an average improvement of +2.41 in code match EM and +2.39 in ID match EM. }
These improvements demonstrate that UER and FSR effectively infer the API information required for code completion, which is crucial for enhancing the accuracy of real-world code completion tasks. 
UER emphasizes code similarity between the code draft and APIs by identifying APIs with naming similarities to those used in the code draft. 
In contrast, FSR focuses on functional similarity, retrieving APIs within the project that provide similar functionality to the code draft. 
Together, these two components complement each other to some extent by addressing different aspects of similarity, thereby enhancing overall performance in inferring the necessary API information for code completion.

\find{
{\bf RQ2:}
In summary, both components of our approach have contributed to its strong performance. When the UER component is added to the minimal baseline, the code match EM increases by an average of +5.65, and the ID match EM increases by an average of +5.88. When the FSR component is added, the code match EM increases by an average of +2.41, and the ID match EM increases by an average of +2.39. These results demonstrate that both components are essential and complementary in improving code and identifier matching performance.
}

\subsection{Complementarity of Our API Inference Method (RQ3)}
\begin{table}
    \centering
    \setlength{\tabcolsep}{4pt}
    \caption{Evaluation results of complementarity of our AIM on DeepSeekCoder-6.7B.}
    \begin{tabular}{cclcccc}
        \toprule[1pt]
        \multirow{2}{*}{Benchmark} & \multirow{2}{*}{Lang} & \multirow{2}{*}{Method} & \multicolumn{2}{c}{Code Match} & \multicolumn{2}{c}{ID Match}\\
        & & & EM & ES & EM & F1\\
         \midrule
         \multirow{12}{*}[-3ex]{ProjBench} & \multirow{6}{*}[-1ex]{Python} 
        & InFile & 18.45 & 45.39 & 23.90 & 41.58 \\
        & & +AIM & \textbf{26.24} & \textbf{50.99} & \textbf{32.25} & \textbf{48.93} \\
         \cmidrule(lr){3-7}
        & & RepoFuse & 23.90 & 49.47 & 30.15 & 47.03 \\
        & & +AIM & \textbf{29.14} & \textbf{52.73} & \textbf{35.20} & \textbf{50.98} \\
         \cmidrule(lr){3-7}
        & & RepoCoder & 24.45 & 50.01 & 30.75 & 48.06 \\
        & & +AIM & \textbf{29.64} & \textbf{53.29} & \textbf{35.65} & \textbf{51.48} \\
         \cmidrule(lr){2-7}
         & \multirow{6}{*}[-1ex]{Java} 
        & InFile & 22.55 & 52.96 & 25.45 & 49.71 \\
        & & +AIM & \textbf{29.15} & \textbf{56.29} & \textbf{33.25} & \textbf{54.74} \\
         \cmidrule(lr){3-7}
        & & RepoFuse & 27.00 & 54.98 & 31.35 & 52.25 \\
        & & +AIM & \textbf{31.80} & \textbf{57.95} & \textbf{36.10} & \textbf{56.64} \\
         \cmidrule(lr){3-7}
        & & RepoCoder & 29.00 & 57.20 & 33.30 & 55.78 \\
        & & +AIM & \textbf{32.35} & \textbf{58.73} & \textbf{36.70} & \textbf{57.79} \\
         \midrule
         \multirow{12}{*}[-3ex]{CCEval} & \multirow{6}{*}[-1ex]{Python} 
        & InFile & 9.08 & 51.32 & 15.87 & 48.01 \\
        & & +AIM & \textbf{25.19} & \textbf{62.20} & \textbf{34.20} & \textbf{61.53} \\
         \cmidrule(lr){3-7}
        & & RepoFuse & 26.79 & 64.11 & 37.00 & 63.52 \\
        & & +AIM & \textbf{34.68} & \textbf{68.81} & \textbf{45.83} & \textbf{69.79} \\
         \cmidrule(lr){3-7}
        & & RepoCoder & 26.87 & 64.55 & 37.60 & 64.39 \\
        & & +AIM & \textbf{34.95} & \textbf{69.14} & \textbf{45.72} & \textbf{70.06} \\
         \cmidrule(lr){2-7}
         & \multirow{6}{*}[-1ex]{Java} 
        & InFile & 10.66 & 52.90 & 17.72 & 51.32 \\
        & & +AIM & \textbf{25.15} & \textbf{60.11} & \textbf{34.32} & \textbf{61.20} \\
         \cmidrule(lr){3-7}
        & & RepoFuse & 22.95 & 57.76 & 31.18 & 57.27 \\
        & & +AIM & \textbf{30.62} & \textbf{62.19} & \textbf{40.16} & \textbf{63.44} \\
         \cmidrule(lr){3-7}
        & & RepoCoder & 25.85 & 61.02 & 35.25 & 61.80 \\
        & & +AIM & \textbf{31.88} & \textbf{63.78} & \textbf{42.17} & \textbf{65.29} \\
         \bottomrule[1pt]
    \end{tabular}
    \label{tab:complementarity}
\end{table}

One of the advantages of our approach is that the UER and FSR can retrieve the necessary internal API information for completion without relying on import statements and can be flexibly integrated into existing methods. To investigate the complementarity of our components, we embed UER and FSR together into various baselines. For convenience, we refer to the combination of UER and FSR as the API inference method (AIM). Specifically, we use the generation results of each baseline as the code draft to retrieve the necessary API information for completion, and then provide the retrieved API information to the model for regeneration. Table \ref{tab:complementarity} shows each method's code match and ID match results. From the table, we can observe that integrating AIM with RAG-based methods improves the model's performance. This indicates that AIM successfully retrieves more comprehensive project knowledge, filling the gaps left by similarity-based retrieval and demonstrating its complementarity. Additionally, we can see that AIM improves the InFile baseline, showing that AIM remains highly effective even when no similar code snippets are present. 
\new{Overall, after integrating our components, existing methods achieve an average improvement of +7.77 in code match EM and +8.50 in ID match EM. }
Previous methods often encounter a knowledge-lack problem because they struggle to locate the necessary internal API information in real-world scenarios. This leads to hallucinations in model predictions. AIM starts from the code draft to infer the internal API information required for completion. It supplements previous methods with additional project-specific knowledge, thereby enhancing their performance. This demonstrates that our UER and FSR components are complementary to existing methods and can be flexibly integrated into them.
\find{
{\bf RQ3:}
In summary, our proposed components (UER and FSR) can be combined with other methods, effectively enhancing the performance of each method. After integrating our components, existing methods achieve an average improvement of +7.77 in code match EM and +8.50 in ID match EM.
}

\begin{table}[ht]
    \centering
    \setlength{\tabcolsep}{10pt}
    \begin{threeparttable}
        \caption{The time costs of different methods. All values are in seconds.}
        \label{tab:time_cost}
        \begin{tabular}{lcc}
            \toprule[1pt]
            Approach & Construction & Inference\\
            \midrule
            RepoFuse & 624.35 & 3.46\\
            RepoCoder & 91.43 & 3.90 \\
            Ours & 348.95 & 4.23\\
            \bottomrule[1pt]
        \end{tabular}
        \begin{tablenotes}
            \footnotesize
            \item ``Construction'' and ``Inference'' refer to the time cost of knowledge base construction and one inference, respectively.
        \end{tablenotes}
    \end{threeparttable}
    
\end{table}

\section{Discussion}
In this section, we discuss the time consumption of our approach, the potential for extension to multi-line code completion of our approach, and semantically equivalent predictions.

\subsection{Time Consumption}
\new{To investigate the time consumption of our approach and other baseline methods, we compare their running time on ProjBench Python, as shown in Table \ref{tab:time_cost}. Specifically, for a given project, the total runtime of the entire method consists of the time for knowledge base construction and the inference time for each task. To ensure reliability and reduce the impact of potential outliers, we conducted the full pipeline 5 times for each method and reported the average runtime.}

\new{
In terms of knowledge base construction, our approach shows a clear advantage over RepoFuse, whose pipeline is more time-consuming because it requires parsing and extracting dependency relationships both between files and within code at a fine-grained level. Although our method incurs more overhead than RepoCoder, which only builds a code snippet database, the additional time is due to constructing an internal API knowledge base using a large model (i.e., Llama3 in our implementation). However, since this process is conducted offline and only requires incremental updates after the initial build, the overhead remains acceptable and practical for real-world scenarios. In summary, while our construction time exceeds RepoCoder’s, it is significantly more efficient than RepoFuse’s, and the offline nature of the process ensures the overhead is reasonable.
}

\new{
Regarding the average inference time, which includes the time of retrieval and target code generation, our approach introduces one additional model call during retrieval compared to RepoCoder. But this only results in 8.46\% time overhead (0.33 seconds) on average.. Compared to RepoFuse, both our approach and RepoCoder include the additional step of code draft generation, leading to a slight delay. However, as discussed in RQ1, this additional cost is justified by the substantial performance improvements. While our approach does introduce some latency due to multiple model inferences, it provides more comprehensive and accurate project-specific knowledge, improving completion accuracy. We also note that our experiments were conducted under limited hardware resources, and both better hardware and more efficient inference acceleration techniques (e.g., faster model runtimes) are expected to further reduce the latency. Therefore, we believe the time overhead is acceptable and will be smaller in realistic, better-optimized deployment settings.
}

\subsection{Extension to Multi-Line Code Completion}
\new{
While our approach is evaluated on project-specific single-line code completion, which is a challenging and important task in practice~\cite{wang2023practitioners}, our approach also holds the potential to be extended to more complex scenarios such as multi-line code completion. For example, in the multi-line setting, the knowledge base constructed by our approach remains fully applicable, as it takes the entire codebase as input to capture internal API information. The primary adaptation lies in the design of the query extraction strategies used in the UER and FSR components. In single-line code completion, UER uses the draft code line as the query line. To adapt UER to multi-line code completion, we can use static analysis and heuristic rules to identify the lines that likely invoke internal APIs from the code draft and use them to construct the query lines. For FSR, which summarizes a code snippet representing a coherent logic unit, we can adapt it by applying code slicing or segmentation techniques~\cite{wang2024hits} to extract multiple semantically meaningful code blocks as queries for retrieval. However, such an extension would necessitate substantial efforts in improving the approach and conducting the evaluation. Thus, we plan to generalize our approach to handle a broader range of code completion scenarios in the future.
}

\subsection{Semantically Equivalent Predictions}
\new{
In our evaluation, we followed prior work~\cite{zhang2023repocoder,ding2024crosscodeeval,ding2022cocomic,liang2024repofuse} and primarily adopted EM as the metric for assessing model performance. While EM provides a straightforward and objective measure, it does not account for predictions that are semantically equivalent to the ground truth but differ in surface form. This may lead to an underestimation of model capabilities. 
}

\begin{figure}[t]
    \includegraphics[width=0.49\textwidth]{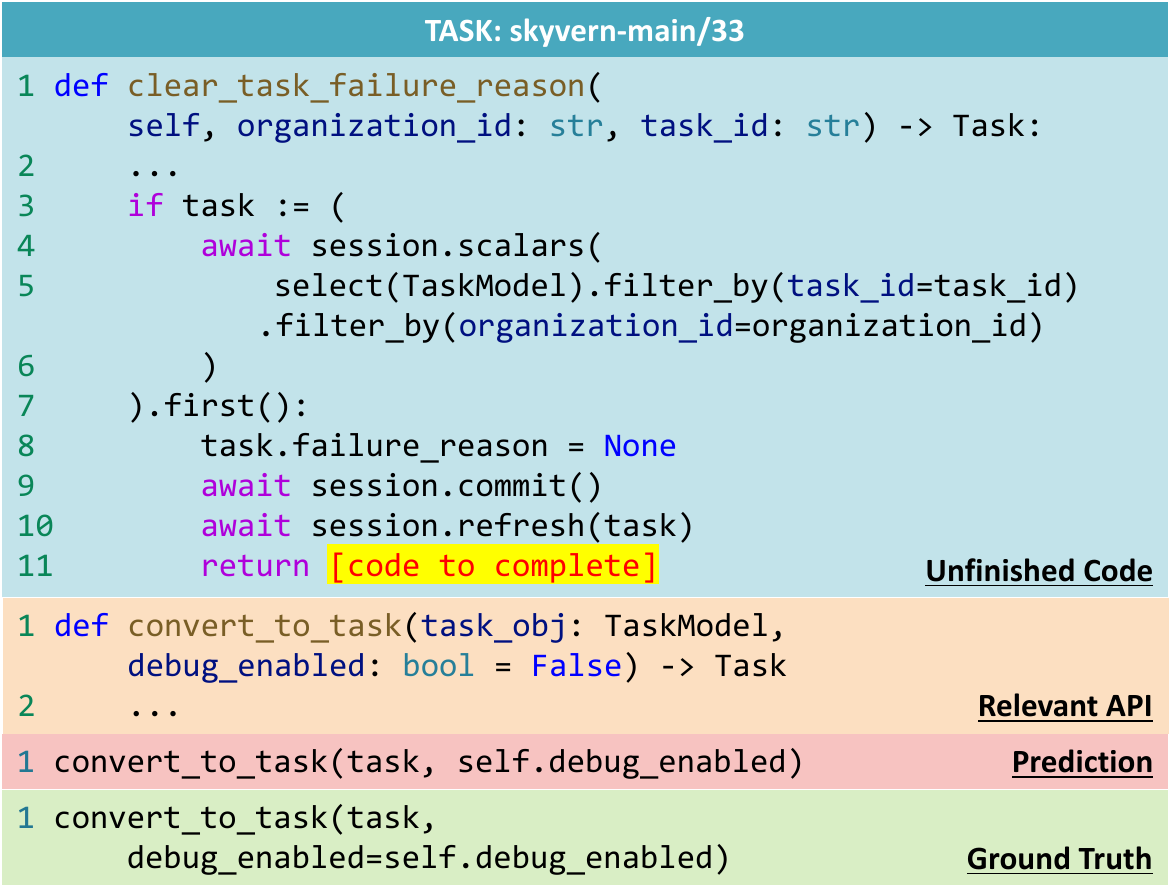}
    \caption{An example in manual evaluation.}
    \label{fig:human_eval}
\end{figure}

\new{
To better understand the extent of this limitation, we manually inspected 50 samples in which the model's prediction failed under EM evaluation. Among these, we identified 3 cases where the prediction was semantically equivalent to the ground truth. Figure \ref{fig:human_eval} illustrates an example with \textit{task\_id skyvern-main/33}. The orange box labeled \textit{Relevant API} contains the API information relevant to the completion. The difference between the \textit{Prediction} and the \textit{Ground Truth} lies in the omission of keyword arguments in the \textit{Prediction}. However, based on the \textit{Relevant API}, it can be observed that these two forms are semantically equivalent in Python.
}

\new{
This preliminary finding highlights the potential value of incorporating semantic-aware evaluation metrics. However, given that the proportion of semantically equivalent predictions among incorrect cases is relatively small and considering the cost of large-scale manual evaluation, we predominantly rely on Exact Match in this study. In future work, we plan to more systematically investigate semantically equivalent predictions, including the integration of automated semantic similarity checks where feasible.
}

\subsection{Evaluation with Latest Commercial LLMs}
\new{
To examine the generalizability of our approach under increasingly capable language models, we conducted an additional evaluation using Claude 3.7 Sonnet~\cite{anthropic2025claude37}, a recent model with a training data cutoff in November 2024. To minimize potential data leakage, we collect 50 single-line code completion tasks from five non-fork open-source projects created after March 2025, ensuring that these examples were unlikely to have been seen during training. The data collection procedure follows the methodology outlined in Section~\ref{sec:dataset_construction}.
}

\new{
We evaluate InFile, RepoCoder, and our approach on 50 samples using Claude 3.7 Sonnet as the underlying model. The results show that our approach correctly completed \textbf{19} out of 50 tasks, significantly outperforming RepoCoder (\textbf{13}) and InFile (\textbf{0}). This substantial gap highlights that, even for advanced LLMs, retrieving and reasoning about project-specific internal API usage remains a non-trivial challenge. Our approach addresses this by leveraging a dedicated internal API knowledge base and retrieval mechanism, which continues to provide substantial benefits even with the latest commercial models. Details of the dataset construction and a representative case study are provided in Appendix B~\cite{appendix}.
}

\new{
While preliminary and limited in scale, this experiment suggests that our approach retains its effectiveness and relevance as foundation models evolve. We consider expanding this line of evaluation to broader scenarios and additional models in future work.
}

\section{Threats to Validity}
\textit{Internal validity.}
A key threat to internal validity lies in potential data leakage, particularly considering the rapid advancement of large language models and their exposure to public code during pretraining. To mitigate this, we carefully constructed the ProjBench benchmark by selecting GitHub repositories created after January 1, 2024, ensuring that their contents are unlikely to have been included in the training data of any existing models. Furthermore, we filtered out forked repositories and explicitly removed import statements from completion contexts to prevent leakage of API references. These steps collectively help isolate the model’s reasoning ability from memorization, enhancing the credibility of the evaluation results. Another concern relates to the accuracy of evaluation metrics. While we primarily report exact match (EM) for both code and identifiers, such metrics may not fully capture cases where the generated code is functionally correct but syntactically different from the ground truth. To address this, we introduced a semantic equivalence analysis as part of our extended evaluation. This analysis examines predictions that achieve the same behavior despite lexical differences, offering a more comprehensive view of model effectiveness. Nevertheless, semantic equivalence remains a challenging problem, and we acknowledge that further development of automatic evaluation tools would enhance robustness.
 
\textit{External validity.}
One limitation in external validity is whether our approach generalizes well to projects outside the benchmark. Although we ensure diversity in ProjBench by selecting large-scale, actively maintained open-source projects across different domains, the selection remains finite and may not cover all development patterns. To strengthen external validity, we additionally evaluated our method on CrossCodeEval, a widely used public benchmark, and observed consistent performance improvements. This supports the claim that our method is not overfitted to a particular dataset and can generalize to broader project types. Our method was primarily designed with Python, a dynamic and widely used language, in mind. However, to test the language generality of our approach, we extended the evaluation to Java, a statically typed language with stricter syntax and type constraints. The results confirm that our method maintains its effectiveness across both dynamic and static languages, demonstrating its flexibility and broader applicability. That said, our approach does rely on language-specific parsing tools (e.g., Tree-sitter) and project metadata, which may require adaptation for less common languages or environments without robust tooling.

\section{Related Work}
\subsection{Large Language Models for Code}
In the past few years, Large Language Models (LLMs) have been extensively researched and applied to code-related tasks, significantly enhancing the programming efficiency of software developers. In the field of code completion, a range of advanced models have emerged, with Codex\cite{chen2021evaluating}, CodeGen\cite{nijkamp2022codegen}, StarCoder\cite{li2023starcoder}, CodeLLaMA\cite{roziere2023code}, Wizardcoder\cite{luo2023wizardcoder}, CodeGeeX\cite{zheng2023codegeex}, and DeepSeekCoder\cite{guo2024deepseek} demonstrating particularly outstanding performance. These models are primarily based on the Transformer decoder-only architecture and trained on large-scale code datasets for next-token prediction tasks, enabling them to provide efficient code completion functionality. These models can be directly used in various Integrated Development Environments (IDEs) to provide real-time code completion services. However, despite their robust performance in most scenarios, these models still exhibit limitations when handling code completion tasks that require cross-file information or internal context dependencies.

\subsection{Project-Specific Code Completion}
In the research field of project-specific code completion, leveraging a broader project context significantly enhances the accuracy and relevance of code completion results. Consequently, the study of project-specific code completion has gained attention, with many efforts~\cite{shrivastava2023repository,shrivastava2023repofusion,ding2022cocomic,phan2024repohyper,zhang2023repocoder,agrawal2023guiding,wang2024teaching,bi2024iterative,liang2024repofuse,lu2022reacc} aimed at capturing and combining project context information. 

RLPG~\cite{shrivastava2023repository} and Repoformer~\cite{shrivastava2023repofusion} train classifiers to identify useful context information. CoCoMIC~\cite{ding2022cocomic} and RepoFusion~\cite{shrivastava2023repofusion} train language models to combine in-file and cross-file contexts and inject knowledge into LLMs. However, these methods depend on labeled data to train their models, making them costly and hard to generalize to unseen projects. 

To obtain project knowledge straightforwardly, many works~\cite{agrawal2023guiding,wang2024teaching,bi2024iterative} combine traditional code tools (e.g., code hint tool, compiler) with LLMs. MGD~\cite{agrawal2023guiding} and TOOLGEN~\cite{wang2024teaching} use suggestions from traditional code hint tools to filter each token generated by the model. ProCoder~\cite{bi2024iterative} uses a compiler to compile the model-generated code and obtains project knowledge based on error messages. The frequent invocation of code tools during inference can limit the efficiency of these methods.

To efficiently obtain complete project dependency information, many works~\cite{liang2024repofuse,phan2024repohyper,ding2022cocomic} use import statements to construct a project context graph. Nevertheless, these works do not consider the situation where, before an internal API is used for the first time within a code file, its corresponding import statement may not exist. The leakage of import statements can cause evaluation results to deviate from practical application. 

To avoid these issues, a series of works~\cite{zhang2023repocoder,lu2022reacc} simply search for related code based on the similarity between code snippets and combine them as a prompt for the LLM, achieving good results. Nonetheless, these methods can not obtain the dependency information required for completion and only rely on highly similar code snippets. When there is little duplicated code in the repository, the generated results will deviate significantly from expectations~\cite{zhang2023repocoder}.

Our method not only retrieves similar code snippets but also obtains the necessary API information for completion without relying on import statements. API information is a supplement to similar code snippets. The combination of these two types of information forms the relatively complete knowledge needed for project-specific code completion. Additionally, our method does not rely on training and treats LLMs as a black box, making it easier and faster to apply in practical development scenarios.

\section{Conclusion and Future Work}
In this paper, we aim to improve project-specific code completion in real-world scenarios by enhancing the understanding and application of internal API information. We first propose a novel method for retrieving internal API information. Instead of relying on import statements, which can be impractical in real-world scenarios, our method first expands internal APIs with usage examples and functional semantic information and then uses a generated code draft to guide the retrieval of the internal API information required for code completion based on these two types of information. Based on this method, we further propose a new approach that combines similar code snippets and API information for better project-specific code completion.
\new{In addition to using the widely used CCEval,}
we craft a new benchmark for project-specific code completion based on real-world Python and Java projects to systematically evaluate our approach. The evaluation results show that our approach outperforms the state-of-the-art baselines by +5.91 and +6.26 in terms of code exact match and identifier exact match, respectively, and demonstrates good generalizability and applicability. In future work, we plan to evaluate the applicability of our framework to more programming languages \new{and more complex scenarios}.

\section*{Acknowledgement}
This research/project is supported by Zhejiang Provincial Natural Science Foundation of China (No.LZ25F020003), National Natural Science Foundation of China (No. 62202420 and No.62302437), Ant Group, and the
National Research Foundation, under its Investigatorship Grant (NRF-NRFI08-2022-0002). Any opinions, findings and conclusions or recommendations expressed in this material are those of the author(s) and do not reflect the views of National Research Foundation, Singapore.

\section*{Data Availability}
The replication package of our method can be found at \url{https://github.com/ZJU-CTAG/InferCom}

\balance
\bibliographystyle{IEEEtran}
\bibliography{main}

\end{document}